\documentclass[twocolumn,amsmath,amssymb,superscriptaddress,nofootinbib,longbibliography,pra]{revtex4-2}

\usepackage{graphicx}
\graphicspath{{images/}}

\usepackage{bm}
\usepackage{dsfont}
\usepackage[dvipsnames]{xcolor}
\usepackage{pstricks}
\usepackage[tight]{subfigure}
\usepackage{verbatim}
\usepackage{units}
\usepackage{natbib}
\usepackage{mathtools}
\usepackage{multirow}
\usepackage{enumitem}
\usepackage{mathrsfs}
\usepackage{leftidx}
\usepackage{xspace}
\usepackage{amsmath}
\usepackage{physics}
\usepackage{tabularx}
\usepackage[normalem]{ulem}  
\usepackage{array} 
\usepackage{tabularx, booktabs}
\newcolumntype{Y}{>{\centering\arraybackslash}X}
\renewcommand{\arraystretch}{1.75}
\usepackage{hyperref}
\hypersetup{colorlinks=true,linktoc=all,linkcolor=blue,breaklinks=true,citecolor=blue,urlcolor=blue}
\usepackage[english]{babel}
\AtBeginDocument{%
\newwrite\bibnotes
\def\bibnotesext{Notes.bib}
\immediate\openout\bibnotes=\jobname\bibnotesext
\immediate\write\bibnotes{@CONTROL{REVTEX42Control}}
\immediate\write\bibnotes{@CONTROL{%
        apsrev42Control,author="08",editor="1",pages="1",title="0",year="1"}}
\if@filesw
\immediate\write\@auxout{\string\citation{apsrev42Control}}%
\fi
}%
\newcommand{\kB}{k_\text{B}}

\newcommand{\Om}{\Omega_\text{mec}}
\newcommand{\omL}{\omega_\text{las}}

\newcommand{\eff}{\text{eff}}
\newcommand{\m}{\text{mec}}
\newcommand{\e}{\text{e}}
\newcommand{\s}{\text{s}}
\renewcommand{\L}{\text{L}}
\newcommand{\R}{\text{R}}
\newcommand{\opt}{\text{opt}}

\newcommand{\Ncav}{\bar{n}_\text{cav}}
\newcommand{\Nm}{\bar{n}_\text{mec}}
\newcommand{\Nc}{|\alpha|^2}
\newcommand{\Nopt}{\bar{n}_\text{opt}}
\newcommand{\Neff}{\bar{n}_\text{fin}}

\newcommand{\Pin}{{P}_\text{las}}

\newcommand{\Gopt}{\Gamma_\text{opt}}

\newcommand{\ain}{\hat{a}_\text{in}}
\newcommand{\dXa}{\delta\hat{X}_a}
\newcommand{\dPa}{\delta\hat{P}_a}
\newcommand{\dXd}{\delta\hat{X}_d}
\newcommand{\dPd}{\delta\hat{P}_d}
\newcommand{\da}{\delta\hat{a}}
\newcommand{\db}{\delta\hat{b}}
\newcommand{\dq}{\delta\hat{q}}
\newcommand{\deltp}{\delta\hat{p}}  
\newcommand{\deltd}{\delta\hat{d}}  
\newcommand{\etaL}{\eta_\text{full}} 
\newcommand{\etaC}{\eta_\text{conv}} 
\newcommand{\mean}[1]{\langle #1 \rangle}

\newcommand{\Tref}[1]{Tab.~\ref{#1}}

\newcommand{\tref}[1]{Tab.~\ref{#1}}
\newcommand{\sref}[1]{Sec.~\ref{#1}}

\begin{document}

\title{Optomechanical cooling with coherent and squeezed light: the thermodynamic cost of opening the heat valve}

\author{Juliette Monsel}
\affiliation{Department of Microtechnology and Nanoscience (MC2), Chalmers University of Technology, S-412 96 G\"oteborg, Sweden\looseness=-1}

\author{Nastaran Dashti}
\affiliation{Department of Microtechnology and Nanoscience (MC2), Chalmers University of Technology, S-412 96 G\"oteborg, Sweden\looseness=-1}

\author{Sushanth Kini Manjeshwar}
\affiliation{Department of Microtechnology and Nanoscience (MC2), Chalmers University of Technology, S-412 96 G\"oteborg, Sweden\looseness=-1}

\author{Jakob Eriksson}
\affiliation{Department of Microtechnology and Nanoscience (MC2), Chalmers University of Technology, S-412 96 G\"oteborg, Sweden\looseness=-1}
\affiliation{University of Gothenburg, S-412 96 G\"oteborg, Sweden}
\author{Henric Ernbrink}
\affiliation{Department of Microtechnology and Nanoscience (MC2), Chalmers University of Technology, S-412 96 G\"oteborg, Sweden\looseness=-1}
\author{Ebba Olsson}
\affiliation{Department of Microtechnology and Nanoscience (MC2), Chalmers University of Technology, S-412 96 G\"oteborg, Sweden\looseness=-1}
\author{Emelie Torneus}
\affiliation{Department of Microtechnology and Nanoscience (MC2), Chalmers University of Technology, S-412 96 G\"oteborg, Sweden\looseness=-1}

\author{Witlef Wieczorek}
\affiliation{Department of Microtechnology and Nanoscience (MC2), Chalmers University of Technology, S-412 96 G\"oteborg, Sweden\looseness=-1}
\author{Janine Splettstoesser}
\affiliation{Department of Microtechnology and Nanoscience (MC2), Chalmers University of Technology, S-412 96 G\"oteborg, Sweden\looseness=-1}

\date{\today}                                                                                                                                                                                                                                                 
\begin{abstract}
Ground-state cooling of mechanical motion by coupling to a driven optical cavity has been demonstrated in various optomechanical systems. In our work, we provide a so far missing thermodynamic performance analysis of optomechanical sideband cooling in terms of a heat valve. As performance quantifiers we examine not only the lowest reachable effective temperature (phonon number), but also the evacuated-heat flow as an equivalent to the cooling power of a standard refrigerator, as well as appropriate thermodynamic efficiencies, which all can  be experimentally inferred from measurements of the cavity output light field. Importantly, in addition to the standard optomechanical setup fed by coherent light, we investigate two recent alternative setups for achieving ground-state cooling: replacing the coherent laser drive by squeezed light or using a cavity with a frequency-dependent (Fano) mirror. We study the dynamics of these setups within and beyond the weak-coupling limit and
give concrete examples based on parameters of existing experimental systems. By applying our thermodynamic framework, we gain detailed insights into these three different optomechanical cooling setups, allowing a comprehensive understanding of the thermodynamic mechanisms at play.
\end{abstract}

\maketitle	
\section{Introduction}

Nano- and micromechanical resonators constitute an excellent platform for exploring thermodynamics on the nanoscale~\cite{Seifert2012Nov,lutz_information_2015,ciliberto_experiments_2017}. Importantly, stochastic and quantum fluctuations can play a major role determining nanomechanical motion~\cite{klaers_squeezed_2017,oconnell_quantum_2010}, enabling tests of stochastic~\cite{Seifert2012Nov} and quantum thermodynamics concepts~\cite{QuThermo} in experiments. In this respect, optomechanical systems~\cite{Aspelmeyer2014Dec} are a pertinent platform as they allow for precise control over the classical and quantum dynamics of nanomechanical motion by using electromagnetic fields. This capability motivated proposals of optomechanics-based work measurement schemes~\cite{Elouard2015May, Dong2015Sep, Monsel2018Nov} or heat engines~\cite{Zhang2014Apr, Zhang2014Aug, gelbwaser-klimovsky_work_2015, Dechant2015May, Mari2015Jul, Bathaee2016Aug, Dechant2017Sep, Bennett2020Oct} and already lead to the experimental quantification of non-equilibrium thermodynamic processes such as irreversible entropy production  \cite{brunelli_experimental_2018,rossi_experimental_2020}.

An ubiquitous thermodynamic process is cooling. In the context of optomechanics, cooling is employed to reduce the entropy of eigenmodes of the mechanical resonator, which is a necessary step for exerting quantum control over mechanical motion and, thus, a crucial requirement for the implementation of any optomechanics-based quantum technology. Furthermore, any optomechanics-based heat engine exploits cooling in its operation cycle. Cooling to the ground state of mechanical motion has been theoretically analyzed (see, e.g., Refs.~\cite{Marquardt2007Aug, wilson-rae_theory_2007, Genes2008Mar, Wilson-Rae2008Sep}) and experimentally realized by direct thermal cooling~\cite{oconnell_quantum_2010}, optomechanical sideband cooling~\cite{Chan2011,teufel_sideband_2011,Delic2020} or feedback cooling~\cite{rossi_measurement-based_2018,magrini_optimal_2020}. In the context of thermodynamics, optomechanical cooling has been examined in the light of entropy production \cite{brunelli_experimental_2018,rossi_experimental_2020} or as a quantum absorption refrigerator \cite{mitchison_realising_2016, Mitchison2018, Naseem2020Jun} fed by thermal light.
However, a detailed analysis of the desired output, such as the reduction in phonon number, related to the \textit{thermodynamic cost} of optomechanical cooling with non-thermal light sources in the steady-state has so far been missing.  

\begin{figure}[htb!]
    \includegraphics[width=\linewidth]{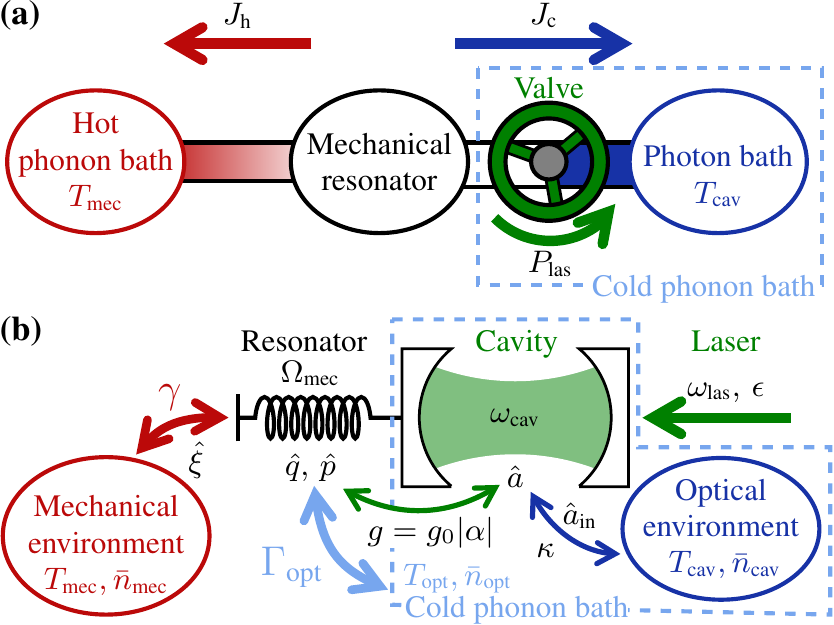}             
    \caption{\label{fig:principle}
        \textbf{(a)} Thermodynamic framework:  a mechanical resonator (in black) is in contact with a hot phonon bath (in red). To cool it down, it is coupled to a photon bath (in blue) whose temperature is much smaller than the relevant photon energies. The coupling can be regarded as opening a valve (in green), allowing heat to flow to the cold bath. The arrows indicate the chosen direction for positive heat flows, but they do not reflect the actual direction of the heat flows in the setup.
        \textbf{(b)} Standard setup for optomechanical cooling: a cavity with a moving-end mirror is driven by a laser. The mechanical resonator is in contact with a thermal phonon reservoir at temperature $T_\m$ and the cavity with a photon reservoir at temperature $T_\text{cav}$. The light blue dashed box represents the effective phonon bath that can be identified in the weak-coupling regime. 
        See Sec.~\ref{sec:II:standard setup} for definitions of the notations. 
    }
\end{figure}

In our work, we present this missing analysis of the thermodynamic performance of different optomechanical sideband cooling schemes, which is of crucial relevance to optimize the performance of future devices.
We therefore provide a comprehensive theoretical framework for sideband cooling beyond previously performed approximations.
We find that this analysis not only yields information about actual performance benchmarks, but also provides insights into the thermodynamic mechanisms at play and the required resources for realizing optomechanical cooling in different experimental setups. 
First, we notice that optomechanical cooling does not correspond to the commonly employed refrigerator framework, in which the work provided to a system makes heat flow out of a cold bath\footnote{See also Appendix \ref{appendix:comp absorption refrigerator} for an overview of the differences between our framework and absorption refrigerators.}. 
The thermodynamic system to be cooled consists of a mechanical resonator (from which heat should be taken away) that is in contact with a hot phonon bath, which leads to Brownian motion of the resonator. On the other hand, there is a photon bath, which is determined by the noise properties of the light field and which appears as cold, since its temperature is much smaller than the energy scale of the photons. This photon bath couples to the mechanical degrees of freedom via a driven optomechanical cavity [see Fig.~\ref{fig:principle}]. We identify that the light field acts as a tunable valve that controls the flow of heat from the mechanical resonator to the cold bath, i.e., the noise of the light field. Therefore, unlike as in usual refrigerators, heat flows from hot to cold, but, importantly, there is a cost associated with maintaining the valve open (via the driving of the cavity). The heat-valve scenario that we suggest here thereby gives a very clear picture of this setting with two heat baths and an additional work input. It is particularly helpful for the understanding of the mechanisms at play when comparing sideband cooling in different optomechanical setups.

We furthermore find it insightful to look at optomechanical cooling as a special example of reservoir engineering \cite{Poyatos1996Dec}: a low temperature \textit{photon} bath is prepared with the help of a laser-irradiated cavity, such that it is transformed into an effective cold \textit{phonon} bath \cite{Marquardt2007Aug}. Engineered reservoirs have shown to be especially appealing to thermodynamics as they allow to explore non-thermal resources~\cite{Brunner2012May, Abah2014Apr, Niedenzu2018Jan, Ghosh2018, Sanchez2019Nov} and study the impact of genuine quantum features, such as squeezing~\cite{Correa2014Feb, Manzano2016May, Agarwalla2017Sep, Manzano2018Oct}, entanglement~\cite{Francica2017Mar} or quantum coherence~\cite{Scully2003Feb, Harrington2019May, Monsel2020Mar} on energy and entropy exchanges. Recently, it was demonstrated that a classically engineered reservoir can be used to beat the classical Carnot bound of heat machines~\cite{klaers_squeezed_2017}. 
Here, the situation is different: we start from a photon bath, in which thermal fluctuations are negligible at optical frequencies, in order to engineer a \textit{thermal} bath of a completely different type, namely an effective phonon bath. In this sense the \textit{valve} that is established is an important ingredient of the engineered bath.
In most cases, the cost to create the engineered bath is not taken into account in the efficiency. 

In our work, we provide and calculate appropriate cooling efficiencies of optomechanical sideband cooling, which account for the cost of the heat valve (i.e.~the bath engineering) and which yield an additional benchmark in optomechanics, besides the resonator's phonon occupation. We then apply our comprehensive thermodynamic analysis of optomechanical cooling to three different setups, which can be used to cool mechanical motion to the ground state. These are: (i) a standard optomechanical setup driven by a coherent laser input [Fig.~\ref{fig:principle}], (ii) a standard optomechanical setup driven by \textit{squeezed-light} [Fig.~\ref{fig:Fano_and_squeezing}(a)], and (iii) an optomechanical cavity that employs a frequency-dependent (Fano) mirror as one of the cavity boundaries and that is driven by a coherent laser [Fig.~\ref{fig:Fano_and_squeezing}(b)]. Importantly, we illustrate our analysis with concrete experimental parameters, which allow us to benchmark different optomechanical platforms against each other. Setup (i)~\cite{Marquardt2007Aug, wilson-rae_theory_2007, Genes2008Mar, Wilson-Rae2008Sep} is widely employed for ground-state cooling provided the optomechanical system is in the resolved-sideband regime~\cite{Chan2011,teufel_sideband_2011}. Setup (ii)~\cite{Asjad2016Nov,Clark2017Jan,Lau2020Mar} is motivated by shaping the quadrature noise of the cold photon bath. This setup can lead to a reduced backaction of the light field onto the mechanical resonator and, thus, to ground-state cooling in a regime, where a coherent drive fails, as we here illustrate. Finally, setup (iii)~\cite{Cernotik2019Jun} results in a Fano-like cavity line shape. This setup aims at suppressing the Stokes process and, thus, the light-induced heating of the mechanical motion originating from the optomechanical interaction itself. Importantly, our theoretical analysis, which employs a Langevin as well as a master equation approach, is valid both for weak and strong optomechanical coupling\footnote{We are referring here to the effective optomechanical coupling $g = g_0\abs{\alpha}$, enhanced by the light field in the cavity, of amplitude $\abs{\alpha}$, see Eq.~(\ref{cooperativity}). On the contrary, the single-photon coupling $g_0$ is assumed to be very small compared to the other frequencies involved.} It thereby extends the validity of previous work~\cite{Asjad2016Nov}.

This paper is structured as follows. In Sec.~\ref{sec:II} we describe the theoretical models we use for each setup, outline the thermodynamic picture that we develop here and present the specific experimental platforms we use as examples. In Sec.~\ref{sec:III} we derive the dynamics of the three setups and apply these results to a thermodynamic analysis of optomechanical cooling illustrated with realistic optomechanics experiments. We conclude in Sec.~\ref{sec:IV}. The appendix contains detailed information on the employed theoretical methods and approximations and thereby allows to straightforwardly follow all calculations performed in this work.

\section{Setups and models} \label{sec:II}

\subsection{Sideband cooling in cavity optomechanical setups}

In the following, we introduce the models for the three different setups that we consider for sideband cooling and start with the standard setup of driving an optomechanical cavity with coherent light.

\subsubsection{Standard setup}\label{sec:II:standard setup}
A typical optomechanical setup consists of a Fabry-Pérot cavity of frequency $\omega_\text{cav}$ with a moving-end mirror of mechanical resonance frequency $\Om$ [see Fig.~\ref{fig:principle}(b)]. The cavity is driven by a laser of frequency $\omL$. It is described by the Hamiltonian~\cite{Aspelmeyer2014Dec}, in the frame rotating at $\omL$,
 \begin{align}\label{H_sys}
   \hat{H}=\,&\hbar \Om \hat{b}^\dagger \hat{b} + \hbar \Delta_0 \hat{a}^\dagger \hat{a} +\hbar (\epsilon \hat{a}^\dagger+ \epsilon^{*} \hat{a})\nonumber\\
   &-\hbar g_0  \hat{a}^\dagger \hat{a} (\hat{b} + \hat{b}^\dagger) ,
\end{align}
where $\Delta_0 = \omega_\text{cav} - \omL$ is the bare detuning between the cavity and the laser. We have denoted $\hat{a}$ the photon-annihilation operator of the cavity and $\hat{b}$ the phonon-annihilation operator. The single-photon optomechanical coupling strength is given by $g_0$ and $\epsilon$ is the drive amplitude. 

The cavity is coupled to a photon and a phonon bath, where
$\kappa$ denotes the optical loss rate and $\gamma$ the mechanical damping rate. The evolution of this system can be described with a set of Langevin equations~\cite{Genes2008Mar}
\begin{subequations}\label{standard:Langevin}
    \begin{align}
        \dot{\hat q}&=\Om \hat p, \\
        \dot{\hat p}&=-\Om \hat q-\gamma \hat p+g_0 \sqrt{2}\hat{a}^\dagger \hat{a}  +\sqrt{\gamma } \hat\xi, \\
        \dot{\hat a}&=-(\kappa+i\Delta_0 ) \hat{a} +ig_0 \sqrt{2}\hat{a} \hat q -i \epsilon +\sqrt{2 \kappa} \ain,
    \end{align}
\end{subequations}
where we have defined the mechanical quadratures $\hat q=\left(\hat{b}+\hat{b}^\dagger  \right)/\sqrt{2}$ and $\hat p=\left(\hat{b}- \hat{b}^\dagger  \right)/i \sqrt{2}$. The thermal noise of the mechanics is given by the operator $\hat{\xi}$, while vacuum noise of the light field is associated with the operator $\ain$. 

We assume in the following that the mechanical quality factor $Q_\m = \Om/\gamma$ is large and that the temperature $T_\m$ of the mechanical environment is high, that is $\hbar\Om \ll \kB T_\m$, see Sec.~\ref{sec:II:experimental setups} for concrete implementations in this regime. We can then make the white-noise approximation for the correlation function for the thermal noise of the mechanics
\begin{equation}
    \mean{\hat \xi(t)\hat \xi(t')}=(2\Nm+1) \delta (t-t'),\label{correlations_xi}
\end{equation}
where $\Nm = (\e^{\hbar\Om/\kB T_\m} - 1)^{-1}$ is the average number of phonons in the mechanical environment.
This approximation is equivalent to the Born-Markov approximation made in the derivation of the master equation~\cite{Wilson-Rae2008Sep} (see Appendix~\ref{appendix:standard:approxs} for a comparison of the different approaches and level of approximations frequently used in optomechanics). In the examples studied in this article (presented in Sec.~\ref{sec:II:experimental setups}), this approximation yields satisfactory results, but we explain how to relax it in Appendix~\ref{appendix:standard:beyond white noise}.

In order to quantify the ratio between the optomechanical coupling and the coupling to the optical and mechanical baths, we use the quantum cooperativity,
\begin{equation}
    C=\frac{2g^2}{\kappa \gamma \Nm}.\label{cooperativity}
\end{equation}
Here, $g=g_0|\alpha|$ is the effective coupling with $|\alpha|$ being the amplitude of the coherent field in the cavity, which is proportional to the driving amplitude of the laser $\epsilon$, see Eq.~\eqref{standard:alpha}. 
\begin{figure}[bt]
    \includegraphics[width=\linewidth]{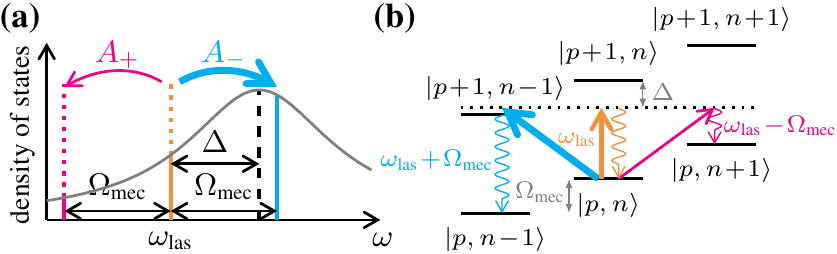}             
    \caption{\label{fig:weak coupling cooling}
        Principle of resolved-sideband cooling in the weak-coupling limit in a Raman-scattering picture. The Stokes and anti-Stokes processes are represented by the magenta and cyan arrows, respectively. 
        \textbf{(a)} Cavity density of states. $\Delta$ is the effective detuning (defined in Sec.~\ref{sec:II:Thermodynamic picture}) between the laser frequency (orange line) and the cavity resonance frequency (dashed black line), which is shifted with respect to $\Delta_0$ by the optomechanical interaction.
        \textbf{(b)} Partial energy diagram of the optomechanical system around the state $\ket{p, n}$ with $p$ photons and $n$ phonons. 
    }
\end{figure}

The laser driving the cavity is modeled as a classical drive, included in the Hamiltonian~\eqref{H_sys}. Its drive amplitude $\epsilon$ is related to the input laser power $\Pin$ by \begin{equation}
    \Pin = \hbar\omL \frac{\abs{\epsilon}^2}{2\kappa}.\label{Pin}
\end{equation}  
The cavity and the resonator are typically at the same temperature, i.e. $T_\m = T_\text{cav}$, however the cavity frequency is typically orders of magnitude larger than the mechanical frequency and $\hbar \omega_\text{cav} \gg \kB T_\text{cav}$ (see Sec.~\ref{sec:II:experimental setups}). Therefore the average number of photons in the optical environment is negligible,  $\Ncav \simeq 0$, and when driving with coherent light, there is only one non-zero correlation function of the cavity input noise, in contrast to the squeezed case, see Eqs.~(\ref{squeezing:correlations_a_in}). This correlation function is
\begin{equation}
     \mean{\ain(t)  \ain^\dagger (t')}= \delta (t-t'). \label{standard:correlations_a_in}
\end{equation}

To sketch the principle of sideband cooling, we consider temporarily the resolved-sideband regime, that is $\kappa \ll \Om$, and a weak optomechanical coupling: $g \ll \kappa, \Om$. In this regime, as illustrated by Fig.~\ref{fig:weak coupling cooling}, three processes can occur: A photon from the laser can (1) be absorbed by the cavity without changing the state of the resonator (in orange), (2) create a photon in the cavity at lower frequency $\omL - \Om$ and a phonon (Stokes process, in magenta), or (3) in combination with a phonon, create a photon at higher frequency $\omL + \Om$ (anti-Stokes process, in cyan). The Stokes process heats up the resonator while the anti-Stokes process cools it down. By choosing a cavity-laser detuning close to the mechanical frequency, the rates of the first two processes can be reduced, allowing  effective cooling~\cite{Schliesser2008May}. Note that the results presented in this article focus on, but are not restricted to, the resolved-sideband regime and they are valid beyond the weak-coupling limit.

\begin{figure}[bt]
    \includegraphics[width=\linewidth]{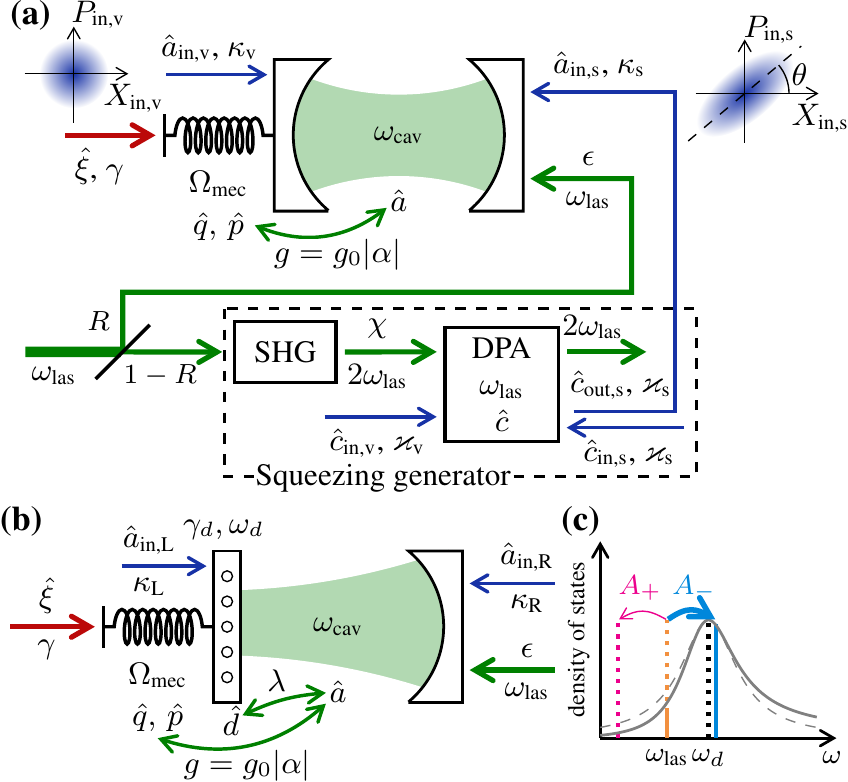}
    \caption{\label{fig:Fano_and_squeezing}
        \textbf{(a)} Squeezed-light setup: We add to the laser driving the cavity a squeezed vacuum field. Since the control of the environment of the cavity is not perfect, there is some residual vacuum noise entering the cavity. The squeezed noise is generated with a degenerate parametric amplifier (DPA) driven at $2\omL$ thanks to a second harmonic generator (SHG). See Sec.~\ref{sec:II:squeezed-light setup} for the definitions of the notations. 
        \textbf{(b)} Fano-mirror setup: The left-hand side mirror of the cavity is replaced by a Fano mirror. Unlike in the canonical setup [Fig.~\ref{fig:principle}(b)], we differentiate the noise coming through the left and right mirrors. See Sec.~\ref{sec:II:Fano-mirror setup} for the definitions of the notations. 
        \textbf{(c)} Cavity density of states (solid gray line) which is asymmetrical in the Fano-mirror setup. The dashed gray line represents the density of states of a standard cavity of equivalent linewidth. 
    }
\end{figure}

\subsubsection{Squeezed-light setup}\label{sec:II:squeezed-light setup}

To improve the cooling scheme presented above, one can use squeezed light instead of coherent light to go beyond the quantum backaction limit~\cite{Asjad2016Nov, Clark2017Jan}. In the model considered here, as depicted in Fig.~\ref{fig:Fano_and_squeezing}(a), this means replacing the vacuum noise input by squeezed noise, generated with a degenerate parametric amplifier (DPA)  (see Ref.~\cite{Gardiner2004}, chapter 10). This is an example of a cascaded open quantum system~\cite{Gardiner1993Apr}, where the output $\hat{c}_\text{out,s}$ of the DPA is used as an input for the cavity, namely $\hat{c}_\text{out,s}(t) = \hat{a}_\text{in,s}(t)$ (we neglect the delay due to propagation, which is irrelevant for our purposes). We consider that the DPA mode, with annihilation operator $\hat{c}$, is resonant with the cavity pump, i.e., at frequency $\omL$. The DPA pump is tuned with the parametric oscillator, i.e., at frequency $2\omL$.

The Langevin equation for the DPA reads~\cite{Asjad2016Nov}
\begin{equation}
    \dot{\hat c}=-\varkappa \hat{c} +\chi \hat{c}^\dagger +\sqrt{2 \varkappa}\hat c_\text{in},
\end{equation}    
where $\hat{c}_\text{in} = \frac{1}{\sqrt{\varkappa}}(\sqrt{\varkappa_\s}\hat{c}_\text{in,s} + \sqrt{\varkappa_\text{v}}\hat{c}_\text{in,v})$ is the total input field, $\varkappa = \varkappa_\s + \varkappa_\text{v}$ the corresponding total loss rate and $\chi =\abs{\chi}\e^{-i2\theta}$ is the squeezing parameter with $\theta$ the squeezing angle [see Fig.~\ref{fig:Fano_and_squeezing}(a)]. The input vacuum noise $\hat{c}_\text{in,v}$ with associated coupling $\varkappa_\text{v}$ corresponds to the uncontrolled losses of the DPA. The desired output from the DPA is then finally obtained from $\hat{c}_\text{out,s}=\sqrt{2\varkappa_\text{s}}\hat{c}-\hat{c}_\text{in,s}$. The Langevin equations for the optomechanical system are still Eqs.~\eqref{standard:Langevin}, but now the total optical input noise reads $\ain = \frac{1}{\sqrt{\kappa}}(\sqrt{\kappa_\s}\hat{c}_\text{out,s} + \sqrt{\kappa_\text{v}}\hat{a}_\text{in,v})$, where $\kappa_\s$ is the coupling to squeezed vacuum noise and $\hat{a}_\text{in,v}$ corresponds to the uncontrolled losses of the cavity with associated coupling $\kappa_\text{v}$~\cite{Asjad2016Nov}. The total loss rate of the cavity is therefore $\kappa = \kappa_\s + \kappa_\text{v}$. In the white-noise approximation, the correlation functions of the input noise read~\cite{Gardiner2004} (see Appendix~\ref{appendix:squeezing:noise})
\begin{subequations}\label{squeezing:correlations_a_in}
    \begin{align}
        \mean{\hat{a}^{\dagger}_\text{in}(t)\ain(t')} &= \pi_\s N_\s\delta(t - t'),\\
        \mean{\ain(t)\ain(t')} &= \pi_\s M_\s \delta(t - t').
    \end{align}
\end{subequations}  
The purity of the squeezing is $\pi_\s = \varkappa_\s \kappa_\s / \varkappa \kappa$ with $\pi_\s = 1$ for pure squeezing and $\pi_\s = 0$ for vacuum noise only. Furthermore, we have  
\begin{subequations}\label{squeeezing:Ns and Ms}
    \begin{align}
    N_\s &= \frac{4 r_\s^2}{(1 - r_\s^2)^2}, \label{squeezing:N_s}\\
    M_\s &= \frac{2 r_\s(1 + r_\s^2)}{(1 - r_\s^2)^2}\e^{-i2\theta}. \label{squeezing:M_s}
\end{align}
\end{subequations}
with the squeezing ratio $r_\s = \abs{\chi}/\varkappa$ and this ratio is such that $0 < r_\s < 1$ so that the white-noise approximation holds. 

\subsubsection{Fano-mirror setup}\label{sec:II:Fano-mirror setup}

An alternative strategy to improve the optomechanical cooling is to make the cavity density of states asymmetric in order to further suppress the detrimental Stokes process [see Fig.~\ref{fig:Fano_and_squeezing}(c)]. This can be achieved by replacing 
the left-hand side mirror of the cavity by a Fano mirror (e.g., a photonic crystal membrane)~\cite{Cernotik2019Jun}.
We then treat separately the cavity input noise $\hat{a}_\text{in,L}$ and $\hat{a}_\text{in,R}$ coming through the left-hand side and right-hand side mirrors, denoting $\kappa_\L$ and $\kappa_\R$ the corresponding loss rates. Both are vacuum noise and their correlation functions obey Eq.~\eqref{standard:correlations_a_in}. The setup is depicted in Fig.~\ref{fig:Fano_and_squeezing}(b) and modeled by adding the internal mode of the Fano mirror to the optomechanical description. We consider an even Fano-mirror mode\footnote{This is the one studied in Ref.~\cite{Cernotik2019Jun}, but it could also be an odd mode, as treated in the Supplementary material of the same article.}, so the setup's Hamiltonian reads~\cite{Cernotik2019Jun}
\begin{equation}
    \hat{H}_\text{Fano} = \hat{H} + \hbar\omega_d \hat{d}^\dagger \hat{d} + \hbar\lambda(\hat{a}^\dagger \hat{d} + \hat{a} \hat{d}^\dagger), \label{H_Fano}
\end{equation}
where $\hat{H}$ is the Hamiltonian of the standard setup, given by Eq.~\eqref{H_sys}. We have denoted $\hat{d}$ the annihilation operator for the considered mode of the Fano mirror, $\omega_d $ its frequency and $\lambda$ the coupling strength between this mode and the cavity mode.

The Langevin equations for such a system read~\cite{Cernotik2019Jun} (in the frame rotating at $\omL$)
\begin{subequations}\label{Fano:Langevin}
    \begin{align}
        \dot{\hat q}=\,&\Om \hat p, \\
        \dot{\hat p}=\,&-\Om \hat q-\gamma \hat p+g_0 \sqrt{2}\hat{a}^\dagger \hat{a}  +\sqrt{\gamma } \hat\xi, \\
        \dot{\hat a}=\,&-(\kappa+i\Delta_0) \hat{a} +ig_0 \sqrt{2}\hat{a} \hat q -i \epsilon - \mathcal{G}\hat{d} \nonumber\\&+\sqrt{2 \kappa_\R}\hat{a}_\text{in,R} +\sqrt{2 \kappa_\L}\hat{a}_\text{in,L},\\
        \dot{\hat d}=\,&-(\gamma_d +i\Delta_d) \hat{d} - \mathcal{G}\hat{a}  +\sqrt{2 \gamma_d }\hat{a}_\text{in,L},
    \end{align}
\end{subequations}
with $\gamma_d$ the coupling strength between the Fano-mirror mode and the left-hand side input, $\Delta_d = \omega_d-\omL$ the detuning with the laser drive and $\kappa = \kappa_\L + \kappa_\R$ the total loss rate of the cavity. We have also defined $\mathcal{G} = i\lambda + \sqrt{\kappa_\L\gamma_d }$. In the model considered here, the coupling between the Fano-mirror mode and the cavity mode is given by  $\lambda = \sqrt{\kappa_0\gamma_d}$~\cite{Cernotik2019Jun}, where $\kappa_0$ is the loss rate that the left-hand side mirror would have if it were frequency independent. For simplicity we take $\kappa_0 = \kappa_\R$.

For the Fano-mirror setup, the total loss rate of the cavity is very large (a lot larger than for the standard setup). Such a large loss rate would place a standard setup in the non-sideband resolved limit. However, this does not reflect the actual linewidth of the cavity. Instead, the presence of the Fano mirror leads to a different effective loss rate, corresponding to the \textit{effective} linewidth of the cavity, $\kappa_\eff$ (see Sec.~\ref{sec:II:experimental setups:Fano}). We therefore use this effective linewidth in the cooperativity
 \begin{equation}
    C\rightarrow\frac{2g^2}{\kappa_\eff \gamma \Nm}\ ,\label{cooperativity_Fano}
\end{equation}
 for this setup, in what follows.

\subsection{Thermodynamic picture} \label{sec:II:Thermodynamic picture}
The thermodynamic system of interest is the mechanical resonator. The phonon number of this mechanical resonator fluctuates, due to its coupling to a mechanical environment, which constitutes a hot phonon bath at temperature $T_\m$. 
The goal of optomechanical (sideband) cooling is to reduce the amount of fluctuations in the resonator, thereby ``cooling it down". 
In order to do so, we have at our disposal a cold photon bath,   in the sense that it contains negligibly few thermal excitations.\footnote{Laser phase noise may be treated as a non-zero temperature bath, see \cite{rabl_phase-noise_2009}.}. However, heat can flow from the fluctuating mechanical resonator to this zero-temperature environment, only if a coupling mechanism between phonons and photons is established. This coupling is provided by the laser that is driving the cavity, thereby constituting a heat valve [see Fig.~\ref{fig:principle} for a sketch of this process]. Hence, the established heat flow from hot to cold comes with a cost, arising from coupling the system to the cold bath, i.e., from keeping the valve opened. 
Note that this \textit{cooling process} is very different from a typical thermodynamic \textit{refrigerator}, where one has the goal to make heat flow from a cold bath to a hot bath by providing work to the system. 

In order to characterize this cooling process, we analyze three thermodynamic quantities, which are complementary to each other and to a certain extent equivalent to those typically quantifying the performance of a thermodynamic cooling device. This is first of all the coldest temperature that can be reached, here represented by the lowest steady-state phonon number of the mechanical resonator. This quantity is experimentally accessible via calibrated optomechanical thermometry or sideband asymmetry determined from the cavity output light field. This is the quantity that is commonly considered when studying optomechanical cooling. Note, however, that the mechanical resonator is a microscopic system and does not constitute a bath itself. By its coupling to the hot mechanical phonon bath and the cold photon bath it is brought into a nonequilibrium state and the lowest reachable  steady-state phonon number can hence only be related to an \textit{effective} temperature. 

Importantly, in our work we complement this study by analyzing two further thermodynamic quantities. Typically, in a refrigerator the  cooling power is analyzed as a performance quantifier, namely the heat flow out of the bath to be cooled. As mentioned above, the mechanical resonator that is intended to be cooled is, however, \textit{not} a bath, and the total heat flow out of or into the resonator is always zero in the steady state. We therefore introduce instead the \textit{evacuated-heat flow} as an equivalent to the cooling power, namely the heat current, $J_\text{c}$, carried by phonons at frequency $\Om$ flowing out of the mechanical resonator and into the effective cold bath.
 We define $J_\text{c} = \hbar\Om I_\text{c}^\text{phonon}$ with the associated phonon flux, $I_\text{c}^\text{phonon}$. The latter quantity is experimentally accessible from the cavity output light field by measuring the amplitude and phase quadrature of the light field, see Appendix~\ref{appendix:standard:flows measurement}.

Finally, we analyze the efficiency of this cooling process.
As described above, maintaining the valve opened in a given position has a cost. The full cost is given by the laser power $\Pin$ required to create the coherent field in the cavity mode. This motivates us to define a cooling efficiency
\begin{equation}
    \etaL = \frac{J_\text{c}}{\Pin}\ ,\label{etaL}
\end{equation}
comparing the evacuated-heat flow (the desired output) with the full laser power (the exploited resource). We define an alternative cooling efficiency later in Eq.~(\ref{etaC}), which, at the cost of neglecting parts of the resource, gives a better picture of the conversion process from phonons to photons. 
This alternative efficiency [Eq.~(\ref{etaC})] can actually be more relevant for device realizations, when the laser light, which was not used in the conversion process, is partly re-used for further operations and, hence, is not lost.
Note that standard refrigerators or heat pumps are often characterized by ``coefficients of performance"; here we refrain from using this terminology and rather use the more generic notion ``(cooling) efficiency", since the desired output of the present process is not  the heat flow into or out of one of the baths, but rather the relevant part of the flux out of the microscopic system attached to them.

In the following, we want to consolidate the above described thermodynamic picture and therefore start with an analysis of the standard setup. As a first step, we linearize the Hamiltonian~\eqref{H_sys} around the system's semiclassical steady state (see Appendix~\ref{appendix:standard:lin} for details)\footnote{This approximation is justified in most optomechanical setups as the optomechanical coupling $g_0$ is typically very small: $g_0 \ll \gamma(2\Nm + 1), \kappa$~\cite{Aspelmeyer2014Dec, Bowen2015Nov}, see \Tref{tab:Params}.}, in which the cavity mode is in the coherent state of amplitude $\alpha$ and the resonator's rest position is $\bar{q}$~\cite{Genes2008Mar}, with
\begin{subequations}\label{standard:average values}
    \begin{align}
        \alpha&= \dfrac{-i\epsilon}{\kappa+i \Delta}, \label{standard:alpha}\\
        \bar{q} &=\sqrt{2}\dfrac{g_0}{\Om} \abs{\alpha}^2.
    \end{align}
\end{subequations} 
We define the effective detuning $\Delta=\Delta_0-g_0 \sqrt{2}\bar{q}$ and choose the phase in $\epsilon$ so that $\alpha=|\alpha|$ is real and positive. We split the operators into their average and fluctuation terms:  $\hat{a}=\alpha + \da$, $\hat{q}=\bar{q} + \dq$, $\hat{p}=\deltp$. The linearized Hamiltonian reads
\begin{align}\label{H_Lin}
    \hat{H}_\text{lin}=\,&\hbar {\Delta} \da^\dagger \da +\hbar \Om \db^\dagger \db\nonumber\\& -\hbar g (\da+\da^\dagger  ) (\db + \db^\dagger) ,
\end{align}
where we have denoted $\db = (\dq + i\deltp)\sqrt{2}$. Indeed, we see from the expression of the effective coupling $g = g_0 \abs{\alpha}$ that the light field in the cavity plays a key role in the cooling process, acting like a valve that allows phonons to flow towards the cavity.

For the sake of obtaining a simple and more usual thermodynamics picture, we---in this section---furthermore resort to the weak-coupling regime ($g \ll \Om, \kappa$).  Tracing out the cavity degrees of freedom, we then find the  master equation~\cite{Wilson-Rae2008Sep} describing the dynamics of the mechanical resonator,
\begin{align}\label{weak-coupling:master_eq}
    \dv{\hat{\rho}_\m}{t}=\,&-i [\Om^\eff\db^\dagger \db, \hat{\rho}_\m]+ (\gamma \Nm + A_+) D[\db^\dagger]\hat{\rho}_\m \nonumber\\ & + (\gamma (\Nm + 1) + A_-)D[\db]\hat{\rho}_\m\ .
\end{align}
Here, $\hat{\rho}_\m$ is the density operator of the resonator, $\Om^\eff$ the effective mechanical frequency (see Appendix~\ref{appendix:standard:solution Langevin}) and $D[\hat{O}]\hat{\rho} = \hat{O}\hat{\rho} \hat{O}^\dagger - \frac{1}{2} \{\hat{O}^\dagger \hat{O}, \hat{\rho}\}$. We have defined the rates
\begin{equation}
    A_\pm = \frac{2g^2\kappa}{\kappa^2 + (\Om \pm {\Delta})^2}\label{Stokes_rates_perturbative},
\end{equation}
which correspond to the Stokes and anti-Stokes processes, respectively.  This is the situation depicted in Fig.~\ref{fig:weak coupling cooling}. 
Equivalently, Eq.~\eqref{weak-coupling:master_eq} can be rewritten in the form
\begin{align}
    \!\!\dv{\hat{\rho}_\m}{t}=\,&-i [\Om^\eff\db^\dagger \db, \hat{\rho}_\m]\nonumber\\
    &+ \Gopt\left( \Nopt D[\db^\dagger] + (\Nopt + 1)D[\db]\right)\hat{\rho}_\m\quad\quad\,\,\,\nonumber\\
    &+ \gamma\left( \Nm D[\db^\dagger] + (\Nm + 1)D[\db]\right)\hat{\rho}_\m.
\end{align}
This equation shows that the driven cavity acts on the resonator like an effective thermal \textit{phonon} bath of associated damping rate $\Gopt \equiv A_- - A_+$ and containing a finite number of phonons $\Nopt = A_+/\Gamma_\opt$. Hence, the cavity and its environment play the role of an effective cold bath (dashed box in Fig.~\ref{fig:principle}(b)), of finite effective temperature $T_\opt$ such that $\Nopt = (\e^{\hbar\Om/\kB T_\opt} - 1)^{-1}$. As a result, the average phonon number associated to the fluctuations of the resonator in the steady state $\Neff = \mean{\db^\dagger \db}_\text{ss}$ can be interpreted as the steady state of a system coupled to two thermal baths~\cite{Aspelmeyer2014Dec},
\begin{equation}
    \Neff = \frac{\gamma\Nm + \Gamma_\opt\Nopt}{\gamma + \Gamma_\opt}.\label{weak-coupling:neff}
\end{equation}
We can see from Eq.~\eqref{Stokes_rates_perturbative} that $\Gamma_\opt$ is proportional to $g^2 = g_0^2 \Nc$. This confirms that the light field in the cavity is like a heat valve allowing more or less heat to flow between the resonator and the cold phonon bath. Note that the effective cold bath is an \textit{engineered bath} and therefore it comes with a cost. This further motivates the above described thermodynamics picture and the resulting definition for the cooling efficiency as given in Eq.~\eqref{etaL}.

In the weak-coupling regime, the evacuated-heat flow is explicitly given by
\begin{align}\label{weak-coupling:Jc}
    J_\text{c} &= \hbar\Om\Gopt(\Neff - \Nopt) \nonumber\\
    &= \hbar\Om\frac{\gamma\Gopt}{\gamma + \Gopt}(\Nm - \Nopt).
\end{align}
Using Eq.~\eqref{standard:average values}, the full efficiency, Eq.~\eqref{etaL}, becomes
\begin{equation}
    \etaL = \frac{\Om 2\kappa}{\omL\alpha^2(\kappa^2 + \Delta^2)}\frac{\gamma\Gopt}{\gamma + \Gopt}(\Nm - \Nopt)\ .
\end{equation}
This weak-coupling thermodynamic picture also gives us further insights into the improvements provided by alternative setups. Squeezing the input noise, as described in Sec.~\ref{sec:II:squeezed-light setup}, allows to reduce the position quadrature noise of the intra-cavity field, which couples to the mechanics. As the cavity transfer function converts part of the input momentum noise into position noise and conversely, it is necessary to rotate the input squeezed state by the adequate angle $\theta$ (given in Sec.~\ref{sec:III:neff:squeezing}) so as to compensate for this effect. This reduction of the intra-cavity position noise can be interpreted as lowering the temperature of the effective cold phonon bath, i.e., $\Nopt$ is smaller while $\Gopt$ remains the same as in the standard setup. On the other hand, with the Fano-mirror setup from Sec.~\ref{sec:II:Fano-mirror setup}, the imbalance between rates of the Stokes and anti-Stokes processes is increased, which results in a change in $\Gamma_\opt$ and a decrease in $\Nopt$. Therefore, the effective cold bath gets colder and its coupling to the resonator is modified.

\subsection{Examples for specific realizations}\label{sec:II:experimental setups}
We now turn to inspect specific optomechanical realizations in order to illustrate our thermodynamic analysis of optomechanical cooling with concrete examples and also to benchmark these realizations against each other. With the evacuated-heat flow and the cooling efficiency, our analysis yields complementary insights beyond the achievable phonon occupation as a benchmark only. To this end, we select four representative optomechanical systems, whose parameters are summarized in \Tref{tab:Params} in the appendix. From here on, we refer to them as system (1) to system (4).

System (1) employs mechanical resonators in the MHz regime and is placed in the sideband-resolved regime. Such systems can be realized, for example, with free-space Fabry-Pérot-based optomechanics~\cite{thompson_strong_2008,groblacher_demonstration_2009}. Importantly, ground-state cooling has already been demonstrated with such a system in the membrane-in-the-middle configuration~\cite{underwood_measurement_2015}. An advantage of the Fabry-Pérot system is its versatility as it can be directly modified to realize a Fano-mirror setup~\cite{Cernotik2019Jun,fitzgerald_cavity_2021}. System (2) is an integrated optomechanics device made of an optomechanical crystal with mechanical motion in the GHz regime that is placed in the sideband-resolved regime. This system is a prominent implementation for realizing optomechanics-based quantum networks with demonstrations of ground-state cooling~\cite{Chan2011} and non-classical state generation~\cite{riedinger_non-classical_2016,riedinger_remote_2018}. Our thermodynamic analysis of this system yields complementary insights beyond its use in quantum networks. System (3) is an optically levitated nanoparticle coupled to an optical cavity. Recently, ground-state cooling of the center-of-mass motion of such a levitated particle has been demonstrated~\cite{Delic2020}. Levitated optomechanics constitutes a pertinent platform for exploring thermodynamic heat engines or stochastic and quantum thermodynamics phenomena. Multiple experiments are already performed along these lines~\cite{gieseler_levitated_2018}. System (4) is based on the membrane-in-the-middle configuration~\cite{thompson_strong_2008}, but in the non-sideband-resolved regime. Recently, feedback-based cooling to the ground state~\cite{Rossi2018} has been demonstrated in such a system. We here show that one can actually achieve ground-state cooling when employing a squeezed-light drive instead of a coherent one, even in the absence of feedback, as evidenced in Refs.~\cite{Asjad2016Nov, Clark2017Jan, Lau2020Mar}. 

We analyze the considered four systems in the context of three different setups: (i) the standard setup, (ii) the squeezed-light setup and (iii) the Fano-mirror setup. The corresponding choice of parameters is shortly discussed in the following.

\subsubsection{Standard setup}

For the standard setup, we assume driving of the optomechanical system with coherent laser light and use the parameters of the four systems as given in \Tref{tab:Params}. In this way, we basically follow the experimental implementations of Refs.\cite{Eichenfield2009Nov, Chan2011,Delic2020,Delic2020_1,Rossi2018}.

\subsubsection{Squeezed-light setup}

In principle, the four systems can also be driven with a squeezed-light source. However, this has so far only been done and used for demonstrating ground-state cooling in microwave optomechanics~\cite{Clark2017Jan}. Squeezed states of light are customarily generated using non-linear optics processes. For example, squeezing levels of 12.7\,dB~\cite{eberle_quantum_2010} and 13\,dB~\cite{schonbeck_13db_2018} at wavelengths of 1064\,nm and 1550\,nm, respectively, have been generated. These wavelengths are also typically used in optomechanics experiments, see \Tref{tab:Params}. Squeezing in these experiments \cite{eberle_quantum_2010,schonbeck_13db_2018} is customarily observed at MHz sideband frequencies, which matches optomechanics implementations with mechanical motion in the MHz regime. Squeezing at GHz sideband frequencies, which is required when employing mechanical motion at GHz frequencies, has been generated at 5\,dB levels~\cite{ast_high-bandwidth_2013,kashiwazaki_continuous-wave_2020,tasker_silicon_2021}. In our analysis, we assume a squeezed-light source with realistic squeezing levels\footnote{The amount of squeezing in dB is calculated as $-10\log_{10}(2\mean{\Delta P^\theta_\text{in,s}})$ with the variance of shot noise being $\frac{1}{2}$ and $\mean{\Delta P^\theta_\text{in,s}}$ the variance of the squeezed quadrature.} of 0.87\,dB, 0.59\,dB, 2.7\,dB and 15.4\,dB for the four systems, respectively. 

\subsubsection{Fano-mirror setup}\label{sec:II:experimental setups:Fano}

Free-space optomechanical devices can be directly adapted to realize a Fano-mirror setup, see, e.g., Refs.~\cite{Cernotik2019Jun,fitzgerald_cavity_2021}. To be able to compare the Fano-mirror setup with the standard setup, we use the fact that the effective linewidth of the cavity with the Fano mirror is $\kappa_\eff = \gamma_d/\zeta_0^2$~\cite{Cernotik2019Jun}, where $\zeta_0$ is the polarizability of the left mirror at frequency $\omega_d$. Since we have chosen $\kappa_\R = \kappa_0 $, $\zeta_0$ is also the (frequency-independent) polarizability of the right mirror. Denoting $\Gamma$ the free spectral range of the cavity that is  given in \tref{tab:Params}, we have $\kappa_\L = 2\Gamma$ and $\kappa_0 = \Gamma/2\zeta_0^2$.
Moreover, choosing to use the optimal value $\gamma_d^* = 4\Om\zeta_0$~\cite{Cernotik2019Jun}, we can determine $\zeta_0$ for each system  such that $\kappa_\eff$ is equal to the system's cavity loss rate.

Optomechanical cavities with Fano mirrors can be realized by using suspended photonic crystal slabs, see for example Refs.~\cite{bui_high-reflectivity_2012,norte_mechanical_2016,chen_high-finesse_2017,gartner_integrated_2018,kini_manjeshwar_suspended_2020}. The parameters of the photonic crystal slab~\cite{fan_analysis_2002} allow engineering of a desired reflectivity at frequency $\omega_d$ and, thus, a corresponding $\kappa_\eff$. Fine tuning of the optical resonance of the photonic crystal mirror can be achieved in post-fabrication, for instance by using thermal tuning~\cite{Peter2005Aug},  strain tuning~\cite{Wong2004Feb} or by depositing~\cite{Yang2007Oct} or etching~\cite{bernard_precision_2016} atomic layers of the suspended photonic crystal. Therefore it is reasonable to assume that $\gamma_d$ close to the optimal value can be achieved in an experimental setup.

\section{Thermodynamic performance}\label{sec:III}

In \sref{sec:II} we have described the cooling process in the three setups and have outlined how to set up a consistent thermodynamic picture. In this section, we analyze the thermodynamic performance of four example realizations and compare their performance for the operation in the three different setups. The parameters used for the four example realizations are given in \tref{tab:Params}. The analysis of the thermodynamic performance includes the lowest reachable phonon number, the evacuated-heat flow as well as the cooling efficiency. In order to obtain all these quantities, we first need to analyze the dynamics of the systems in the three setups.

\subsection{Dynamics}

The first step to the study of the systems' dynamics in each of the three setups is to linearize the Langevin equations, as we have previously done in Sec.~\ref{sec:II:standard setup} for the standard setup. Here we present a short overview for all three setups. We then derive the evolution of the second-order moments, and in particular of the photon and phonon numbers.

\subsubsection{Standard setup}\label{sec:III:dynamics:standard}
Applying the linearization introduced in Sec.~\ref{sec:II:Thermodynamic picture} to the Langevin equations for the standard setup, Eq.~\eqref{standard:Langevin}, we obtain
\begin{subequations}\label{standard:Langevin lin}
    \begin{align}
        \delta \dot{\hat{q}}&=\Om  \deltp, \\
        \delta\dot{\hat{p}}&=-\Om  \dq-\gamma  \deltp +g\sqrt{2}  (\da^\dagger + \da)  +\sqrt{\gamma } \hat{\xi}, \\
        \delta \dot{\hat{a}}&=-(\kappa+i\Delta )  \da +ig\sqrt{2}  \dq +\sqrt{2 \kappa} \ain.
    \end{align}
\end{subequations}
From these equations and the noise correlation functions, given by Eqs.~\eqref{correlations_xi} and~\eqref{standard:correlations_a_in}, we derive the evolution of the photon and phonon numbers associated to the fluctuations\footnote{Note that the environment contribution in Eq.~\eqref{standard:evolution phonon number} has an extra term in $\mean{\db^2 + \db^{\dagger 2}}$ compared to the one in Ref.~\cite{Wilson-Rae2008Sep} because Ref.~\cite{Wilson-Rae2008Sep} used a different noise model. Both models give very similar results in the high mechanical quality factor regime considered here. See Appendix~\ref{appendix:standard:approxs} for a discussion of the differences between the two approaches.} 
\begin{subequations}\label{standard:evolution photon and phonon numbers}
    \begin{align}
    \!\dv{\mean{\da^\dagger \da}}{t} =\,
    & i g \left(\mean{\da^\dagger (\db^\dagger +  \db)} - \mean{\da (\db^\dagger+ \db)}\right)\nonumber\\
    & - 2\kappa\mean{\da^\dagger \da}, \label{standard:evolution photon number}\\ 
    \!\dv{ \mean{\db^\dagger \db}}{t} =\,
    & ig\left(\mean{( \da +  \da^\dagger)\db^\dagger} - \mean{(\da^\dagger+\da)\db }\right)\nonumber\\
    &\!\!\!+ \gamma \left(\Nm - \mean{\db^\dagger \db}  + \frac{\mean{\db^2 + \db^{\dagger 2}}}{2}\right).\!\label{standard:evolution phonon number}
    \end{align}
\end{subequations}
Details about the derivation and the evolution equations of the other second-order moments can be found in Appendix~\ref{appendix:standard:Lyapunov}.
For each of the above equations, two contributions stand out: the optomechanical one, proportional to $g$, and the one from the optical and mechanical environment, proportional to $\kappa$ and $\gamma$, respectively. When one neglects the non-resonant processes $\da\db$ and $\da^\dagger\db^\dagger$ (rotating wave approximation, valid for small coupling $g$, see Appendix~\ref{appendix:standard:approxs}), the optomechanical contribution corresponds to the scattering picture from Fig.~\ref{fig:weak coupling cooling}(b), with heating from the Stokes process and cooling from the anti-Stokes process.

\subsubsection{Squeezed-light setup}

In the case of the squeezed-light setup, described in Sec.~\ref{sec:II:squeezed-light setup}, the DPA and cavity are only coupled through the noise. Therefore, the average fields in the cavity and resonator are still given by Eqs.~\eqref{standard:average values} and the linearized Langevin equations are the same as Eqs.~\eqref{standard:Langevin lin}. In contrast, the cavity input noise correlation functions are given by Eqs.~\eqref{squeezing:correlations_a_in}. As a consequence, the photon number evolution becomes
\begin{align}
    \dv{\mean{\da^\dagger \da}}{t} =\,
    & i g \left(\mean{\da^\dagger (\db^\dagger +  \db)} - \mean{\da (\db^\dagger+ \db)}\right)\nonumber\\
    & +2\kappa(\pi_\s N_\s - \mean{\da^\dagger \da} )\label{squeezing:evolution photon number},
\end{align}
while the evolution equation of the phonon number is unchanged (Eq.~\eqref{standard:evolution phonon number}).
Further changes to the evolution of the second-order moments are detailed in Appendix~\ref{appendix:squeezing:Lyapunov}.

\subsubsection{Fano-mirror setup}

Like in Sec.~\ref{sec:II:Thermodynamic picture}, we linearize the Langevin equations~\eqref{Fano:Langevin} by splitting operators into their average terms and fluctuation terms. But in this case, we have one extra operator, $\hat{d}=\delta + \deltd$, which modifies the semiclassical steady state into (see Appendix~\ref{appendix:Fano:lin} for details)
\begin{subequations}\label{Fano:average values}
    \begin{align}
    \alpha&=\frac{-i\epsilon}{\kappa+i \Delta - \mathcal{G}^2/(\gamma_d  + i\Delta_d )}\label{Fano:alpha} \\
    \bar{q} &=\sqrt{2}\dfrac{g_0}{\Om} \abs{\alpha}^2,\\
    \delta&=\frac{-\mathcal{G}}{\gamma_d  + i\Delta_d } \alpha.
\end{align}
\end{subequations}
The linearized Langevin equations read
\begin{subequations}\label{Fano:Langevin lin}
    \begin{align}
        \delta \dot{\hat{q}}=\,&\Om  \deltp, \\
        \delta\dot{\hat{p}}=\,&-\Om  \dq-\gamma  \deltp +g\sqrt{2}  (\da^\dagger + \da)  +\sqrt{\gamma } \hat{\xi}, \\
        \delta \dot{\hat{a}}=\,&-(\kappa+i\Delta )  \da +ig\sqrt{2}  \dq -\mathcal{G}\deltd \nonumber\\ &+\sqrt{2 \kappa_\L} \hat{a}_\text{in,L}+\sqrt{2 \kappa_0} \hat{a}_\text{in,R},\\
        \delta \dot{\hat{d}}=\,&-(\gamma_d +i\Delta_d  )  \deltd -\mathcal{G}\da +\sqrt{2 \gamma_d } \hat{a}_\text{in,L}.
    \end{align}
\end{subequations}
The free parameter $\Delta_d$ of the Fano-mirror mode was found to be constrained by the cavity mode and the optical loss rates by  $\Delta_d = \Delta - 2\sqrt{\kappa_0\kappa_\L}$~\cite{Cernotik2019Jun}.

From these equations and the noise correlation functions, given by Eqs.~\eqref{correlations_xi} and \eqref{standard:correlations_a_in}, we derive the evolution of the photon number (see Appendix~\ref{appendix:Fano:Lyapunov}):
\begin{align}
    \dv{\mean{\da^\dagger \da}}{t} =\,
    & i g \left(\mean{\da^\dagger (\db^\dagger +  \db)} - \mean{\da (\db^\dagger+ \db)}\right) \nonumber\\
    & -2 \Re{\mathcal{G}\mean{\deltd \da^\dagger}}  -2\kappa\mean{\da^\dagger \da} \label{Fano:evolution photon number},
\end{align}
The evolution equation of the phonon number is unchanged (Eq.~\eqref{standard:evolution phonon number}).

\subsection{Steady-state phonon number}

The first performance indicator we consider is the steady-state phonon number $\Neff = \mean{\db^\dagger \db}_\text{ss}$, which is the standard quantity that is used to characterize sideband cooling and experimentally accessible via calibrated noise thermometry or sideband asymmetry. In a thermodynamic picture, this quantity would correspond to the lowest temperature that can be reached by a refrigerator.   
Indeed, the average phonon number in the fluctuations can be related to $T_\eff$, the \textit{effective} temperature of the resonator, by
\begin{equation}
   \Neff = \left(\e^{\hbar\Om/\kB T_\eff} - 1\right)^{-1}.
\end{equation}
 Since the position of the resonator has been shifted (see Sec.~\ref{sec:II:Thermodynamic picture}), we actually have the mean stationary phonon number $\mean{\hat{b}^\dagger \hat{b}}_\text{ss} = (g/\Om)^2\Nc + \Neff$. 
 
 In the following, we describe how to obtain the steady-state phonon number, $\Neff$, in the three setups and discuss some of its general properties. We then discuss differences between the different setups and example realizations in the end of this section.

\subsubsection{Standard setup} 

For the standard setup, $\Neff$ can be found analytically, by solving the equations for the second-order moments for the steady state. However, the obtained expression is rather complex and we therefore present it in Appendix~\ref{appendix:standard:Lyapunov}, see also Refs.~\cite{Genes2008Mar,Wilson-Rae2008Sep}. Instead, in order to illustrate our results, we here present plots, see Fig.~\ref{fig:neff}, using the realistic experimental parameters from Tab.~\ref{tab:Params}. The only free parameter in the standard setup is the detuning $\Delta$. In the resolved-sideband regime, $\kappa \ll \Om$, the lowest phonon number is obtained for $\Delta \simeq \Om$, as illustrated in Fig.~\ref{fig:weak coupling cooling}. Note however, that the optimal detuning is exactly given by the mechanical frequency only in the weak-coupling approximation. All the plots for the standard setup are done at $\Delta = \Om$, except for the experimental system (4) (see Tab.~\ref{tab:Params}), where we use $\Delta = \kappa$. System (4) is not in the resolved-sideband regime and the cavity linewidth is larger than the mechanical frequency. As a consequence, if $\Delta = \Om$, the Stokes and anti-Stokes rates are almost equal, not leading to effective cooling. The best cooling is obtained when the difference between the two rates is the largest, that is on the slope of the cavity density of states, for a detuning close to $\kappa$. 

\subsubsection{Squeezed-light setup} \label{sec:III:neff:squeezing}
As in the standard case, we obtain an analytical expression of the steady-state phonon number (see Appendix~\ref{appendix:squeezing:Lyapunov}) for the squeezed-light setup.
Again, the lowest phonon number is obtained for $\Delta \simeq \Om$. However, we have three additional parameters: the squeezing purity $\pi_\s$, the squeezing ratio $r_\s$ and the squeezing angle $\theta$ (see Sec.~\ref{sec:II:squeezed-light setup}). As expected, the best results with respect to the lowest reachable phonon number are obtained for pure squeezing, i.e. $\pi_\s = 1$. Besides, there exists an optimal value of the squeezing angle $\theta^*$ that minimizes the phonon number.
Interestingly, this angle does not depend on the other squeezing parameters (see Eq.~\eqref{squeezing:optimal angle} in Appendix~\ref{appendix:squeezing:optimization} for the full expression) and, in the limit of weak coupling and neglecting the mechanical damping rate, it is given by~\cite{Asjad2016Nov}
\begin{equation}
    \tan(2\theta^*) = \frac{2  {\Delta} {\kappa}}{{\Delta}^{2} - {\Om}^{2} - {\kappa}^{2}}.
\end{equation}
Therefore it is possible to determine this optimal angle from the optomechanical setup's parameters. There exists also an optimal value $r^*_\s$ of the squeezing ratio which depends on the squeezing angle (see Appendix~\ref{appendix:squeezing:optimization} for details).
All the plots for this setup are done using the parameters from Tab.~\ref{tab:Params}, with $\Delta = \Om$ (except for system (4) for which $\Delta = \kappa$), $\pi_\s = 1$ and the  values of $\theta$ and $r_\s$, which are optimal for reducing the steady-state phonon number.

Going to the limit of weak coupling and neglecting the mechanical damping rate, we derive the phonon number in the effective cold bath (see Sec.~\ref{sec:II:Thermodynamic picture}) 
\begin{eqnarray}
    \Nopt^\text{squ} & = & \Nopt\left( 1 + \frac{2\pi_\s N_\s(\Delta^2 + \Om^2 + \kappa^2)}{\kappa^2 + (\Delta - \Om)^2} \right.\\
&&   \left. -2 \pi_\s \abs{M_\s}\sqrt{\frac{\kappa^2 + (\Delta + \Om)^2}{\kappa^2 + (\Delta - \Om)^2}}\cos(2\theta - 2\theta^*) \right),\nonumber
\end{eqnarray} 
where $\Nopt$ is the phonon number in the effective cold bath in the absence of squeezing, in agreement with the results from~\cite{Asjad2016Nov}. This explicitly shows that, with the appropriate squeezing parameters, we can get a \textit{colder} effective cold phonon bath than with the standard setup, thus allowing to reach a lower $\Neff$. In this limit, the optimal squeezing ratio $r_\s^*$ is such that 
\begin{equation}
    \frac{\abs{M_\s(r_\s^*)}}{N_\s(r_\s^*)} = \sqrt{\frac{\kappa^2 + (\Delta + \Om)^2}{\kappa^2 + (\Delta - \Om)^2}}
\end{equation}
and then $\Nopt^\text{squ} = \Nopt(1 - \pi_\s)$~\cite{Asjad2016Nov}. Therefore, in this ideal case and with perfect squeezing, the lowest attainable temperature of the effective cold phonon bath is zero.

\subsubsection{Fano-mirror setup} 
In this setup, the cavity density of states is centered around $\omega_d$ \cite{Cernotik2019Jun} [see Fig.~\ref{fig:Fano_and_squeezing}(c)], therefore the lowest phonon number can be obtained for $\Delta_d \simeq \Om$~\cite{Cernotik2019Jun}. As explained in Sec.~\ref{sec:II:experimental setups:Fano}, Fano mirrors with the desired loss rate $\gamma_d$ can be fabricated by engineering the photonic crystal lattice parameters. It is therefore realistic to use the loss rate $\gamma_d^*$, for which the phonon number gets minimized, see Tab.~\ref{tab:Params} for the numerical values, and $\Delta_d = \Om$ in the plots. Once again, to obtain the best cooling, we use a different detuning for system (4): $\Delta_d = 2.5\Om$ (see Appendix \ref{appendix:Fano:non-sideband-resolved}).
We solved the equations for the second-order moments in order to obtain $\Neff$  
numerically (see Appendix~\ref{appendix:Fano:Lyapunov}).

\subsubsection{Discussion}

\begin{figure}[tb!]
    \includegraphics[width=\linewidth]{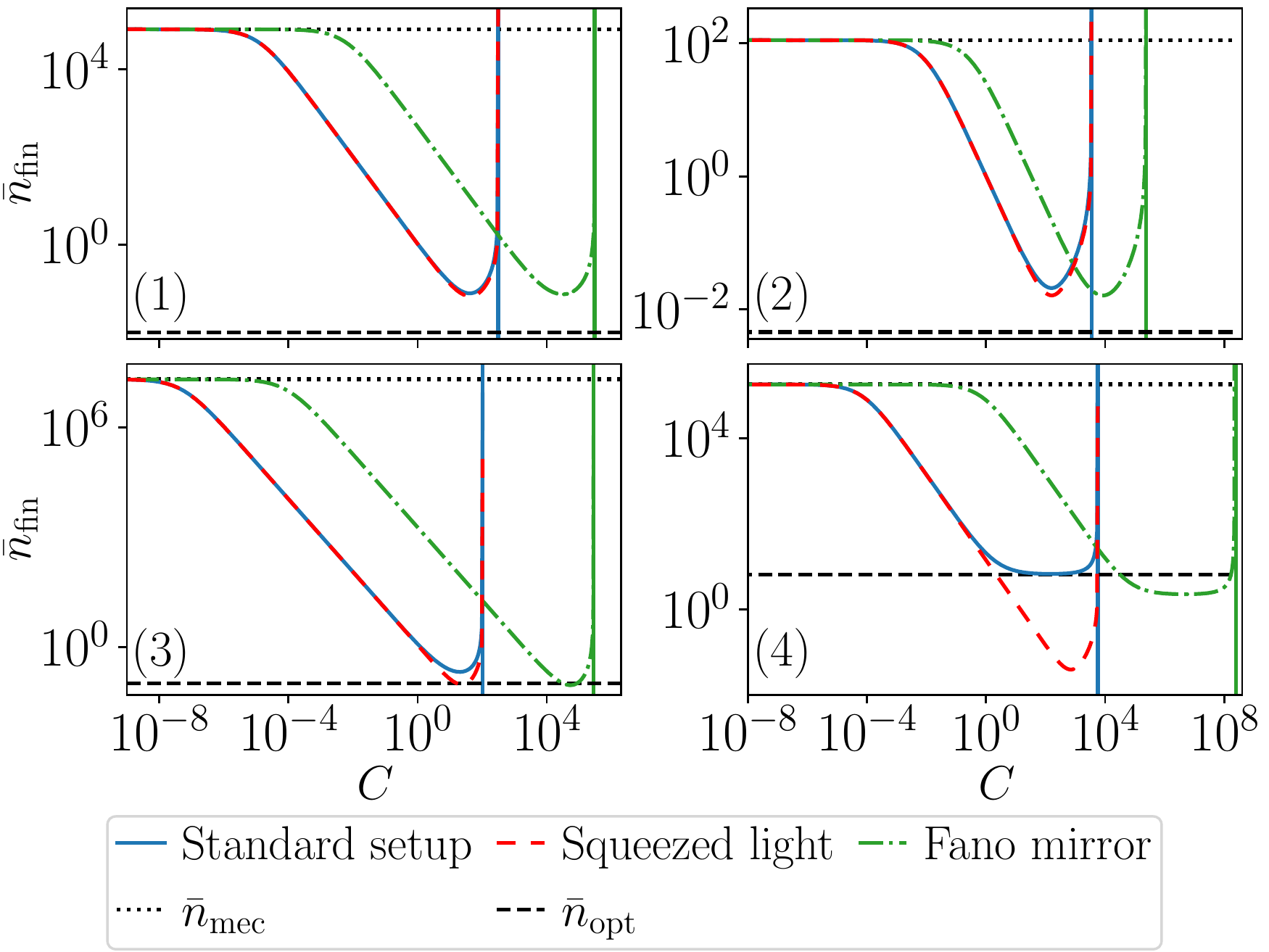}
    \caption{\label{fig:neff}
        Steady-state phonon number as a function of cooperativity for the three setups (indicated by different lines in each plot). The dotted and dashed black lines respectively indicate the average phonon number in the hot mechanical bath and in the effective cold bath (weak-coupling limit) for the standard setup. The vertical solid lines indicate the cooperativities at which the standard and squeezed-light setups (in blue) and Fano-mirror setup (in green) become unstable (see Appendices \ref{appendix:standard:lin} and \ref{appendix:Fano:lin}).The number in each sub-figure corresponds to the system number as given in Tab.~\ref{tab:Params}.
    }
\end{figure}
\begin{figure}[b!]
    \includegraphics[width=\linewidth]{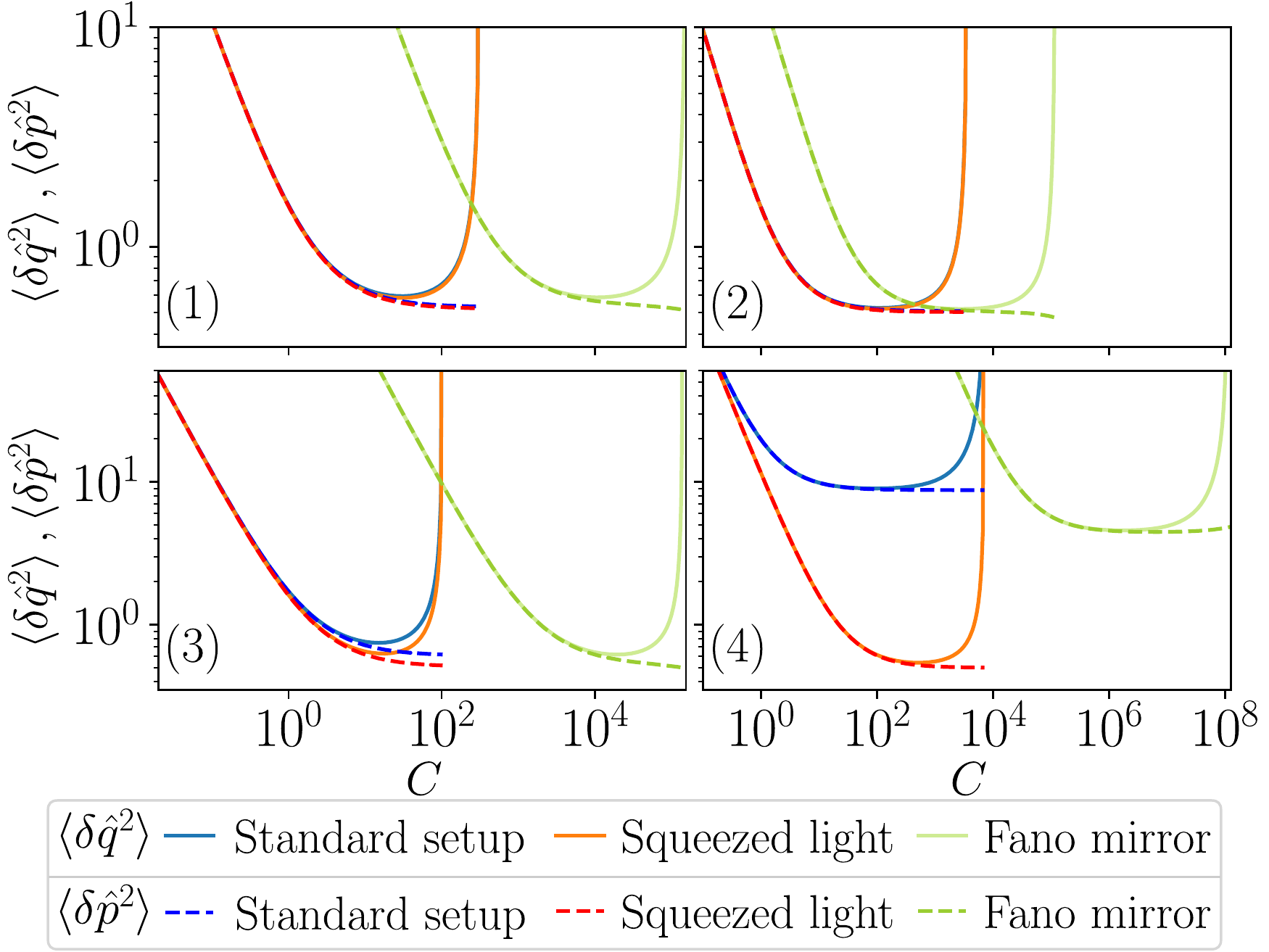}
    \caption{\label{fig:quadratures}
        Fluctuations of the mechanical position (solid lines) and momentum (dashed lines) quadratures as a function of cooperativity for the three setups: standard setup in blue (dark gray), squeezed-light setup in red and orange (medium gray) and Fano-mirror setup in green (light gray). The number in each sub-figure corresponds to the system number as given in Tab.~\ref{tab:Params}. For systems (1) to (3) the curves for the standard and squeezed-light setups are almost identical.
    }
\end{figure}

We have plotted $\Neff$ as a function of cooperativity for each setup and for each system from Tab.~\ref{tab:Params} in Fig.~\ref{fig:neff}.  For all four example systems in all three cooling setups, we notice  that at low cooperativity, the thermal noise prevails over the optomechanical coupling and $\Neff \simeq \Nm$. Then, when the cooperativity increases, the optomechanical coupling allows to reach lower phonon numbers, in most cases down to the ground state of the mechanical resonator (namely $\Neff<1$). The minimum is  reached, when the optomechanical coupling becomes so large that $\Neff$ is governed by the occupation of the effective cold phonon bath, namely $\Neff\approx \Nopt$ (see Eq.~\eqref{weak-coupling:neff} obtained in the weak-coupling limit). Finally, when further increasing the cooperativity, the coupling becomes so strong that the optomechanical back-action increases the fluctuations of the resonator and the phonon number increases as well\footnote{In the strong coupling limit, lower phonon numbers can be obtained in an instantaneous instead of the steady state, see Ref.~\cite{Liu2014May}.}. This is the case when non-resonant processes of the type $\delta\hat{a}\delta\hat{b}$ and $\delta\hat{a}^\dagger\delta\hat{b}^\dagger$ start to play a role. This behavior due to back-action is not captured by the weak-coupling approximation (Eq.~\eqref{weak-coupling:neff}, see also Appendix~\ref{appendix:standard:approxs}). Eventually, the phonon number diverges\footnote{
The system is no longer stable, see Appendices \ref{appendix:standard:lin} and \ref{appendix:standard:Lyapunov}.}. Interestingly, Fig.~\ref{fig:quadratures} shows that the divergence of $\Neff = (\mean{\dq^2} + \mean{\deltp^2}  -1)/2$ fully comes from the position fluctuations while the momentum fluctuations keep decreasing with cooperativity.

There are, however, also differences between the different systems and between the different setups. 
Comparing the different setups, we find that both the driving with squeezed light as well as the use of a Fano mirror allow to reach lower phonon numbers than the standard setup, as expected. Interestingly, the steady-state phonon number is however only little decreased with respect to the standard setup. For systems (1) to (3), the difference is hardly visible on the log-scale used in Fig.~\ref{fig:neff}; while the absolute difference in the phonon number is indeed small, 
the relative difference with respect to the standard setup is quite significant, as visible in Tab.~\ref{tab:rel diff x}.

\begin{table}[tb]
    \setlength{\tabcolsep}{10pt}
    \begin{tabular}{c|cccc}
        \hline
         System&  (1) &  (2) &  (3) &  (4) \\ \hline\hline
        $x^\text{squeezed}$ & 14.4\% & 22.5\% & 54.8\% &99.4\% \\\hline
        $x^\text{Fano}$ &  5.0\%& 22.7\%& 56.5\% & 66.4\%\\ \hline
    \end{tabular}
    \caption{\label{tab:rel diff x}
       Relative difference between the minimum phonon number $\Neff$ reached with the standard and the squeezed-light/Fano-mirror setup, $x^\text{squeezed/Fano}=\left(\Neff^\text{standard}-\Neff^\text{squeezed/Fano}\right)/\Neff^\text{standard}$.
    }
\end{table}

Furthermore, the dips in the curves for the Fano-mirror setup are always shifted to larger cooperativity values, which means that a larger optomechanical coupling is required to cool the resonator. This can be reached by increasing the power of the laser driving. The shift can be understood by looking at the weak-coupling picture. For the systems we consider, the coupling $\Gopt$ to the effective cold phonon bath, at a given $C$, is weaker in the Fano-mirror setup than in the standard setup (see Appendix~\ref{appendix:Fano:weak-coupling}).

For all systems in the resolved-sideband regime (the first three in Tab.~\ref{tab:Params}), all setups enable ground-state cooling. While the overall behavior seems very similar for systems (1) to (3), the initial phonon number has very different orders of magnitude in the different systems, meaning that also the overall reduction in order to reach the ground state is not the same. In contrast system (4), which is not in the resolved-sideband regime, does not allow for ground state cooling when using a coherent drive\footnote{This would require feedback cooling (see for instance~\cite{Genes2008Mar}), as was done in the experiment using this setup~\cite{Rossi2018}.}. However, driving with squeezed light  reduces $\Neff$ below one. Numerical values for the lowest reachable phonon numbers in all setups and systems are given in \Tref{tab:Params}.

\subsection{Evacuated-heat flow}

Traditionally when studying refrigerators, one looks at the cooling power, namely the heat that is extracted from the object to be cooled. Here, heat is extracted from the mechanical resonator by phonons flowing into the engineered, cold environment, thereby compensating the heat flow from the hot mechanical bath. The evacuated-heat flow, constituting an equivalent to the cooling power here, hence  corresponds to the phonon flow into this cold environment times the frequency of the mechanical resonator, namely the energy that each of these phonons carries away, $J_\text{c} = \hbar\Om I_\text{c}^\text{phonon}$. This quantity can be inferred from measurements of the cavity output light field
, see Appendix~\ref{appendix:standard:flows measurement} for details.

The flux of phonons from the resonator to the cavity, $I_\text{c}^\text{phonon}$, can be identified in the evolution of the phonon number (Eq.~\eqref{standard:evolution phonon number}) as
\begin{equation}
   I_\text{c}^\text{phonon} = ig\mean{(\da + \da^\dagger)(\db - \db^\dagger)}.\label{cooling_power}
\end{equation}
Note that this expression is very similar to the contribution $\mu_b$ of the entropy production rate analyzed in Ref.~\cite{brunelli_experimental_2018} with $I_\text{c}^\text{phonon} = -\mu_b(\Nm +\frac{1}{2})$ in the steady state. The flux $-I_\text{c}^\text{phonon}$ is then equal to the flow of phonons going into the mechanical environment, which is an irreversible process that contributes to the entropy production rate and, thus, is reflected in $\mu_b$.
\begin{figure}[htb!]
    \includegraphics[width=\linewidth]{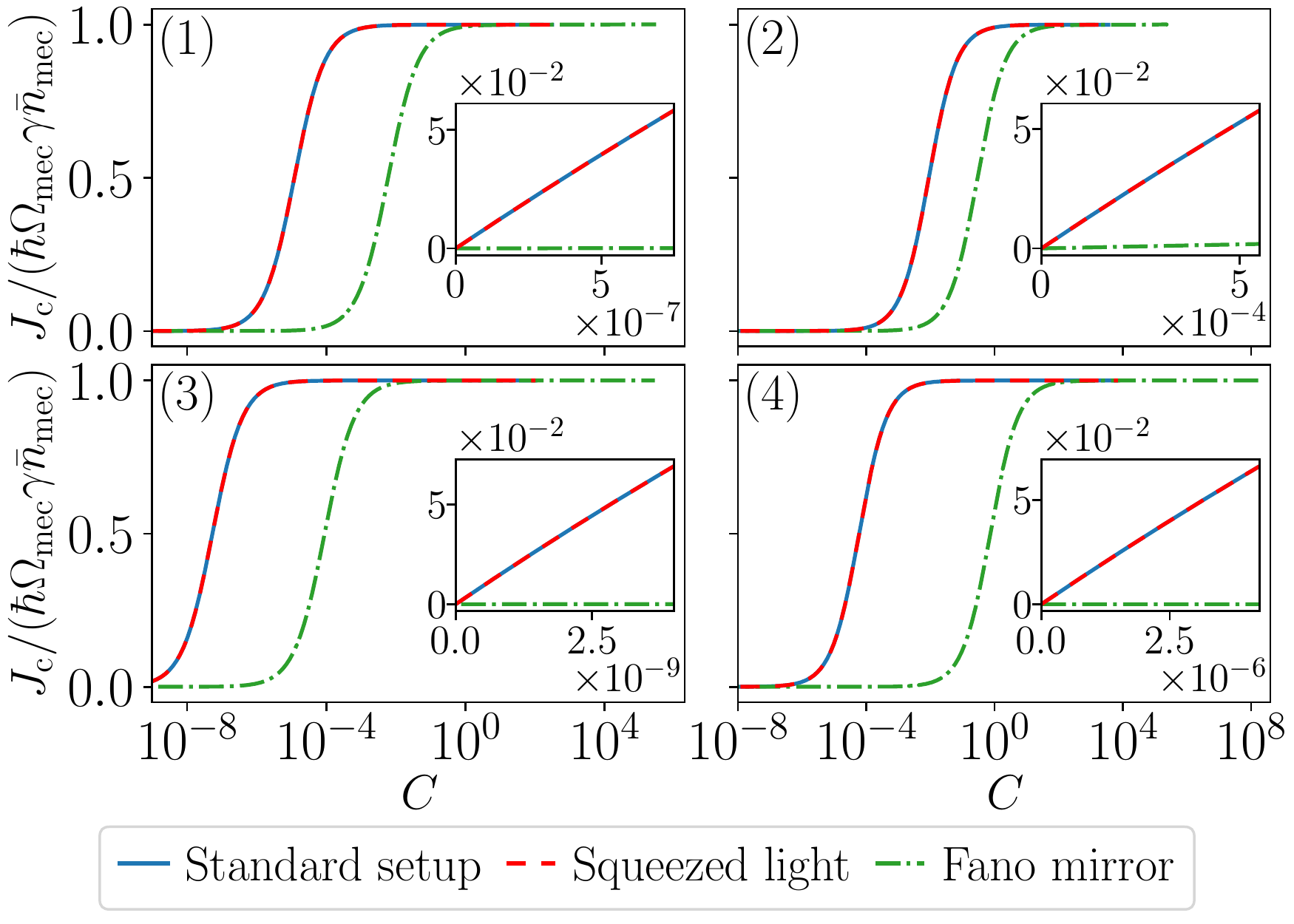}
    \caption{\label{fig:cooling_power}
        Evacuated-heat flow as a function of cooperativity for the three setups (indicated by different lines in each plot). The insets show the evacuated-heat flow at low cooperativity with a linear scale. The number in each sub-figure corresponds to the system number in the parameter Tab.~\ref{tab:Params}. 
    }
\end{figure}

We obtain the steady-state evacuated-heat flow in the three setups and for the four different systems and show  $J_\text{c}$ in Fig.~\ref{fig:cooling_power} as a function of the cooperativity, up to those values, where $\Neff$ diverges. 
For small cooperativity values, the evacuated-heat flow is linear (visible in the insets) but vanishingly small, namely when the optomechanical coupling is small. The result for the evacuated-heat flow from the weak-coupling approximation (Eq.~\eqref{weak-coupling:Jc}), $J_\text{c}\approx\hbar\Om\Gopt\Nm$,  clearly shows that the phonon flow for small $C$ is constrained by the small coupling to the effective cold phonon bath.
It then rapidly increases at cooperativity values at which $\Gopt\approx\gamma$. This is the same parameter range for which the steady-state phonon number $\Neff$ starts to differ from the value imposed by the hot environment, $\Nm$. The evacuated-heat flow then saturates at a constant value, $J_\text{c}\approx\hbar\Om\gamma\Nm$,  namely when the coupling to the effective optical bath is so large that the flow of phonons into the cold bath is only constrained by
the inflow of phonons from the hot mechanical bath. This is the maximum amount of (phononic) heat that can flow through the system, as long as the effective optical phonon bath is colder than the mechanical bath. Note that this saturation happens at a much smaller cooperativity than the one at which the minimum $\Neff$ is reached. Indeed, the bottleneck of the heat flow through the system is reached when $\Gopt$ exceeds $\gamma$; however, equilibration of the resonator at some $\Neff$ is governed by the effective cold bath only when the exchange rate $\Gopt\Nopt$ exceeds $\gamma\Nm$.
Importantly, the evacuated-heat flow is \textit{not} affected by the backaction effect leading to an increase of $\Neff$ at large cooperativity. This means that the opposite heat flow from the non-resonant terms $\delta\hat{a}\delta\hat{b}$ and $\delta\hat{a}^\dagger\delta\hat{b}^\dagger$ exactly counterbalance each other, see  Appendix~\ref{appendix:standard:heat flows} and the plot therein. The strong hybridization between cavity photons and phonons in the limit of large cooperativity can hence be understood as leading to a direct steady-state heat flow between mechanical bath and light field, with less cooling effect on the local mechanical sub-system.  Note that this crucial difference between the behaviors of the lowest reachable phonon number and the evacuated heat flow clearly show the complementarity of these two performance quantifiers.

The described qualitative behavior is globally the same in all setups and for all studied systems. 
The threshold, which is needed to significantly cool down the resonator, is however higher for the Fano mirror, since, as discussed previously, higher cooperativities are needed to establish a $\Gopt$ that is comparable to the one of the standard and the squeezed-light setup. 
The standard setup and the setup fed by squeezed light only differ by small numerical values. This means that the evacuated-heat flow is not influenced by the noise properties of the input light. Indeed, in the weak-coupling regime that governs the physics of the evacuated-heat flow, squeezing only enters via $\Nopt$, see Eq.~(\ref{fig:weak coupling cooling}), which is negligibly small both in the standard and the squeezing setup.

\subsection{Efficiencies}
The squeezed-light and Fano-mirror setups allow to reach lower phonon numbers than the standard setup, without altering the saturation value of the evacuated-heat flow. In this section, we aim to understand if these cooling schemes are more  or less efficient, namely whether they require extra resources to evacuate the same amount of heat and reach resonator states close to the ground state. We therefore look at the cooling efficiency $\etaL$ defined as the evacuated-heat flow divided by the laser power (Eq.~\eqref{etaL}) as well as a complementary cooling efficiency $\etaC$, characterizing the efficiency of the phonon-photon conversion process.   

\subsubsection{Full cooling efficiency}

\begin{figure}[b!]
    \includegraphics[width=\linewidth]{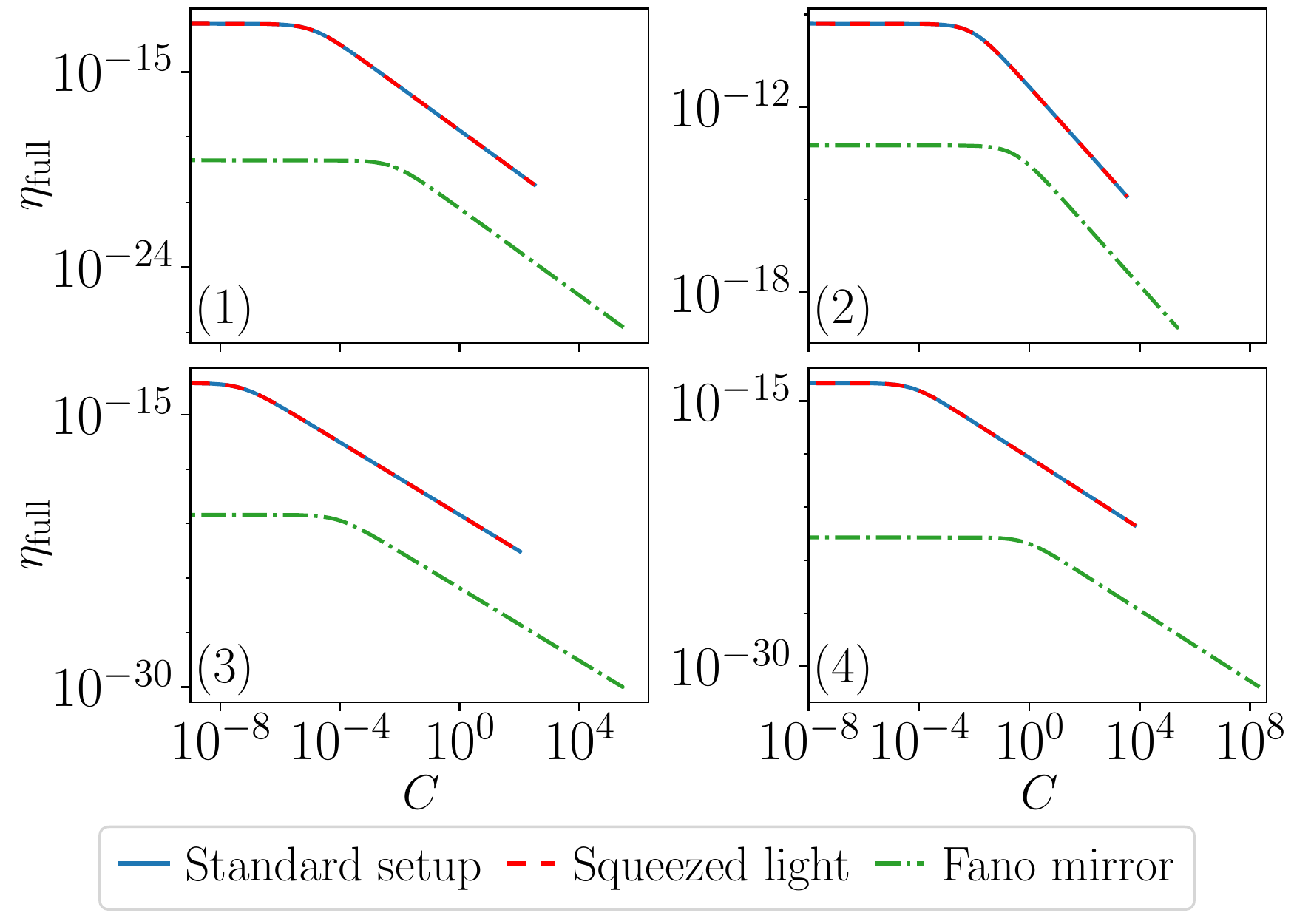}
    \caption{\label{fig:etaL}
        Full cooling efficiency $\etaL$ as a function of cooperativity for the three setups (indicated by different lines in each plot). The number in each sub-figure corresponds to the system number in parameter Tab.~\ref{tab:Params}. Note that the full cooling efficiency of the squeezed-light setup differs from the plotted one by a factor $R$ as given in Eq.~\eqref{eq:etaprime}.
    }
\end{figure}

In order to obtain the full cooling efficiency for all setups and systems, we insert the expression for the laser power, Eq.~\eqref{Pin}, and for the evacuated-heat flow, Eq.~\eqref{cooling_power}, into the full efficiency, $\etaL = J_\text{c}/\Pin$. The drive amplitude $\epsilon$ of the laser, entering the expression for the laser power is determined from the steady-state amplitude of the field in the cavity,  $|\alpha| = g/g_0$, using Eq.~\eqref{standard:alpha} in the standard and squeezed-light setups, and Eq.~\eqref{Fano:alpha} for the Fano-mirror setup.
The resulting full cooling efficiency is plotted in Fig.~\ref{fig:etaL} as a function of the cooperativity.

For all setups and systems, the cooling efficiency  is constant until some cooperativity threshold, namely the one at which the evacuated-heat flow displays a crossover to a plateau. The reason for this is that both the evacuated-heat flow as well as the laser power are linear in $C$ for small cooperativities. The small magnitude of the full cooling efficiency derives from the low phonon energy compared to the energy of the laser photons and the weak optomechanical single-photon coupling $g_0$. At cooperativities at which the evacuated-heat flow sets in, the full cooling efficiency decreases rapidly. We chose to cut the curves at the cooperativities, in which the phonon number $\Neff$ diverges.

The full cooling efficiencies for the three setups have similar features, but differ by their magnitude. The largest efficiency is the one of the standard setup, since it requires the least laser-power input. For the squeezed-light setup, only the reflected fraction $R$ of the input laser power is actually used for driving the cavity, while the remaining transmitted part feeds the DPA, generating the squeezing, see Fig.~\ref{fig:Fano_and_squeezing}(a). Therefore, the actual efficiency, $\etaL'$, is smaller than the plotted one, $\etaL$, by a factor $R$,
\begin{equation}\label{eq:etaprime}
    \etaL'=\frac{J_\text{c}}{ \Pin/R}=R\etaL
\end{equation}
The factor $R$ by which the full cooling efficiency of the squeezed-light setup is reduced with respect to the one of the standard setup hence depends on the cost of the squeezing generation.  We refrain from giving a specific value for $R$, because quantifying this cost requires to fix the details of the experimental setup used to generate the squeezed vacuum. With Eq.~\eqref{eq:etaprime}, we provide an instruction on how to include the cost of squeezing in a concrete experimental realization.

The full cooling efficiency of the Fano-mirror setup is orders of magnitude smaller than the one of the standard (and squeezed-light) setup. Indeed, as visible in Eqs.~\eqref{Fano:average values}, for the Fano-mirror setup, the laser field not only creates the steady-state field in the cavity but also the one in the mirror mode. So more laser power is required to reach the same value of $\alpha$, and therefore the same cooperativity. Thus, $\etaL$ illustrates how much extra resources are required to cool down the resonator using a Fano mirror. 

Overall, reaching lower phonon numbers, either with squeezed light (as one has to account for the factor $R<1$ taking into account the cost to create squeezed light) or a Fano mirror, is hence more costly and reduces the efficiency of the cooling process.

Comparing the efficiencies of the example realizations of different systems, we note that the full cooling efficiency of the non-resolved sideband system (4) is similar to systems (1) and (3), even though the  phonon number $\Neff$ of system (4) is larger than the one for the standard and Fano-mirror setups. This motivates us to define in the following section another cooling efficiency that is more sensitive to whether the system is in the resolved-sideband regime.

\begin{figure}[tb!]
    \includegraphics[width=\linewidth]{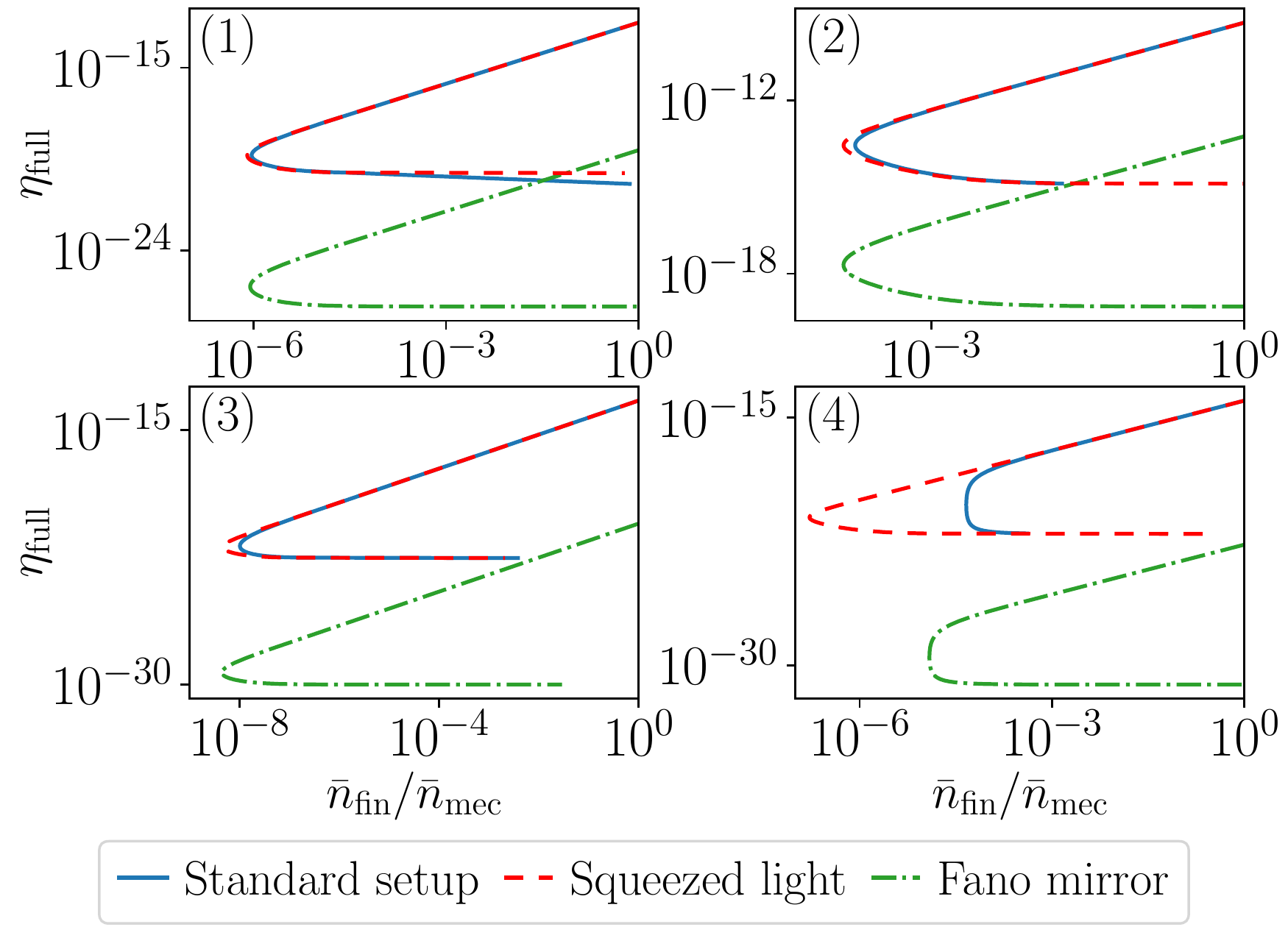}
    \caption{\label{fig:loop_etaL}
        Full cooling efficiency $\etaL$ as a function of the effective phonon number for the three setups (indicated by different lines in each plot). The number in each sub-figure corresponds to the system number in parameter Tab.~\ref{tab:Params}. Note that the full cooling efficiency of the squeezed-light setup differs from the plotted one by a factor $R$ as given in Eq.~\eqref{eq:etaprime}.
    }
\end{figure}

As is commonly the case in thermodynamics, there is a trade-off between optimizing the efficiency and the desired output. While in a standard refrigerator, this trade-off is typically analyzed between the efficiency and the cooling power, we here identify the lowest reachable phonon number as the relevant quantity which should be optimized, possibly together with the efficiency of the heat evacuation process. We illustrate this trade-off in Fig.~\ref{fig:loop_etaL}. Indeed, $\etaL$ starts to decrease at a cooperativity orders of magnitude lower than the one at which $\Neff$ reaches its minimum value.

\subsubsection{Conversion efficiency}

The full cooling efficiency, $\etaL$, discussed above compares the evacuated-heat flow to the \textit{full} laser power input needed to set up the cooling mechanism. While this yields a rather complete benefit/cost ratio, it gives less insight about the cooling efficiency of the optomechanical conversion process from phononic to photonic heat. Note however, that the light leaking out of the cavity could in principle partly be reused; thus, this part would further be available as a resource. In this subsection, we therefore introduce an alternative cooling efficiency that we call \textit{conversion efficiency}.

When looking at the microscopic process, it is of interest to consider the photons involved in the optomechanical interaction as the resource. We therefore start by identifying the flux of photons interacting with the mechanics 
\begin{equation}
    I^\text{photon} = ig\left(\mean{\da^\dagger (\db^\dagger +  \db)} - \mean{\da (\db^\dagger+ \db)}\right). \label{photon flow}
\end{equation}
We have obtained this quantity by analyzing the evolution of the photon number, Eq.~\eqref{standard:evolution photon number}. Note that this expression is very similar to the contribution $\mu_a$ of the entropy production rate analyzed in Ref.~\cite{brunelli_experimental_2018} with $I^\text{photon} = \mu_a/2$ in the steady state. Again, this is not surprising as $I^\text{photon}$ describes the flux of photons that ultimately leak out of the cavity to the optical environment, which is an irreversible process that contributes to the entropy production rate. Further, also $I^\text{photon}$ can be inferred from measurements of the output cavity light field, for details see Appendix~\ref{appendix:standard:flows measurement}.

With this, the conversion efficiency can be defined as the evacuated-heat flow (phonons flowing out of the resonator into the cold bath) with respect to the flow of photons that actually interact with the mechanics, 
\begin{align}
    \etaC & = \frac{I_\text{c}^\text{phonon}}{I^{\text{photon}}}\ .\label{etaC-0}
\end{align}
Using Eqs.~\eqref{cooling_power} and \eqref{photon flow}, we obtain
\begin{align}
    \etaC 
    &=\frac{\mean{\da^\dagger\db-  \da\db^\dagger } - \mean{\da^\dagger\db^\dagger -  \da\db}}{ \mean{\da^\dagger\db-  \da\db^\dagger } + \mean{\da^\dagger\db^\dagger -  \da\db}}\ .\label{etaC}
\end{align}
In sideband cooling schemes, the detuning of the laser drive is close to resonance with the beam-splitter terms, $\da^\dagger\db$ and $\da\db^\dagger$, while the two-mode-squeezing terms, $\da\db$ and $\da^\dagger\db^\dagger$, are non-resonant. The latter are detrimental to the cooling, i.e. $\mean{\da^\dagger\db^\dagger -  \da\db} > 0$ (a plot of the beam-splitter and two-mode-squeezing contributions to the heat flow is included in Appendix \ref{appendix:standard:heat flows}), leading to $\etaC < 1$. However when the beam-splitter terms are perfectly resonant, that is for $\Delta=\Om$ in the resolved-sideband regime the two-mode-squeezing contributions become negligible in the weak-coupling limit and the conversion efficiency goes to 1.

\begin{figure}[tb!]
    \includegraphics[width=\linewidth]{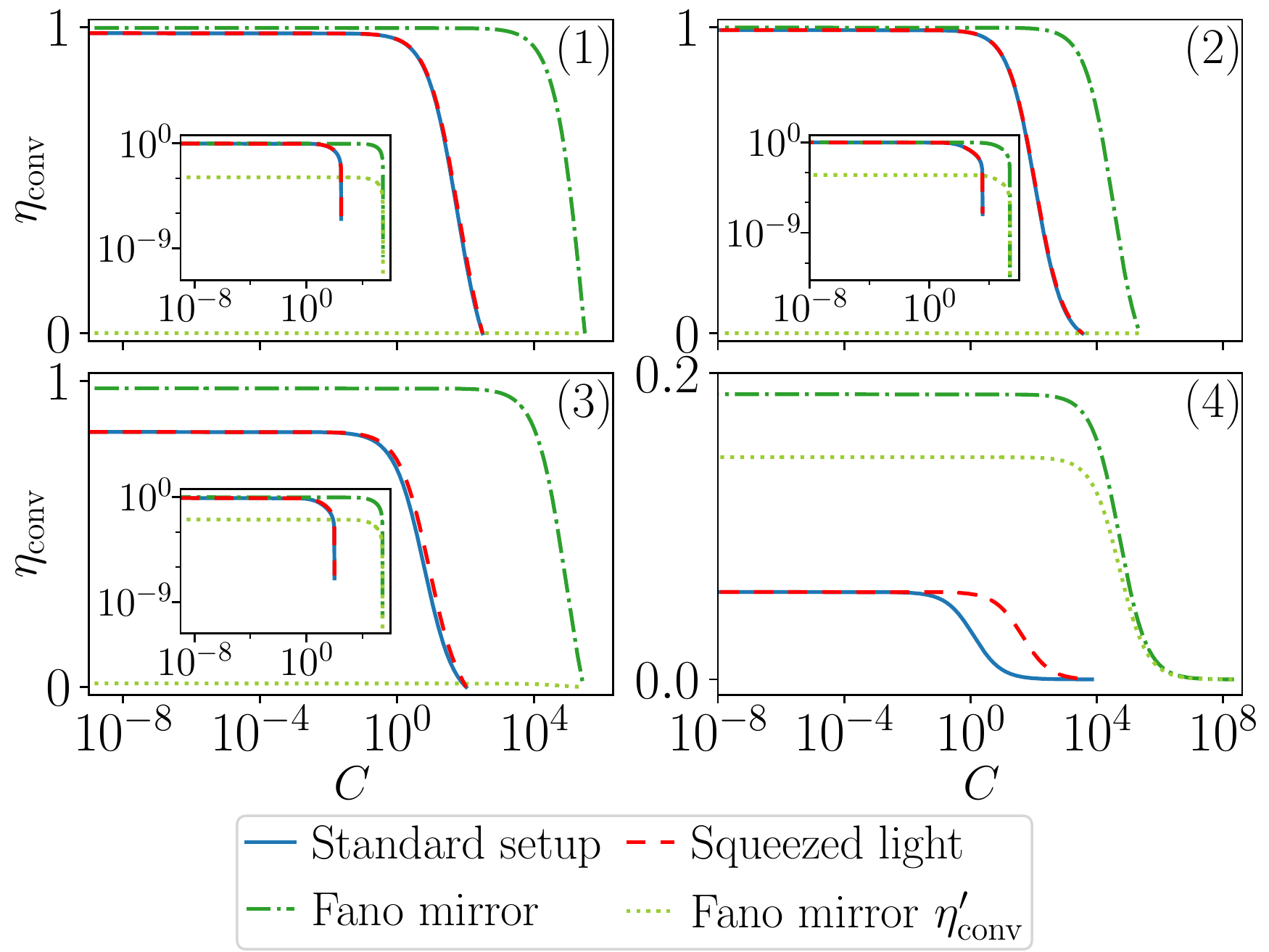}
    \caption{\label{fig:etaC}
        Conversion efficiency $\etaC$ as well as $\etaC'$ as a function of cooperativity for the three setups (indicated by different lines in each plot). The insets show the same plots but in log-log scale. The number in each sub-figure corresponds to the system number in parameter Tab.~\ref{tab:Params}. The insets show sections of the main plots as log-log plots.
    }
\end{figure}
Note that in the Fano-mirror setup, there is an extra term in the evolution of the photon number (Eq.~\eqref{Fano:evolution photon number}) due to the added mirror mode. This extra term does neither interact with the mechanics nor is it leaking out of the cavity. If interested in this extra cost of the interaction between the cavity and the Fano-mirror mode as well, one can take this term into account as one of the resources in the efficiency, which then gives
\begin{equation}\label{etaC_Fano}
    \!\!\!\etaC' = \frac{g\Im( \mean{\db^\dagger (\da^\dagger+\da)})}{ g\Im( \mean{\da (\db^\dagger+\db)}) - \Re(\mathcal{G}\mean{\deltd \da^\dagger}) }.\!
\end{equation}

Figure~\ref{fig:etaC} shows the conversion efficiency, $\etaC$, as a function of cooperativity for all setups and systems. 
One first of all observes that the conversion efficiency is by orders of magnitude larger than the full cooling efficiency and can even reach values up to 1. This means that the largest reduction of resources provided by the laser comes from the re-usable laser power, which is accounted for in $\etaC$ but not in $\etaL$. 
At the same time, it shows that it is the conversion efficiency, $\etaC$, which provides the deepest insights into the fundamental physics of our cooling process.

Interestingly, while the full efficiency $\etaL$ rapidly decreases at those cooperativity values at which the evacuated-heat flow starts to be considerably large, the conversion efficiency, $\etaC$ stays close to maximal up to those cooperativity values, at which the phonon number, $\Neff$ approaches its minimum.

The conversion efficiency is overall the largest for the Fano-mirror setup, while it is very similar for the squeezed-light and standard setups. This similarity between standard and squeezed-light setup reflects that the effect of squeezing is mainly a reduction of the temperature of the cold effective phonon bath. Instead, the presence of the Fano mirror improves the conversion process. 
 However, the modified conversion efficiency of the Fano-mirror setup, $\etaC'$,  which takes into account the cost of the interaction between the cavity and Fano-mirror modes (light green dotted line), is reduced by up to several orders of magnitude. That can be seen best in the insets of the plot panels of Fig.~\ref{fig:etaC}.  
\begin{figure}[tb!]
 \includegraphics[width=\linewidth]{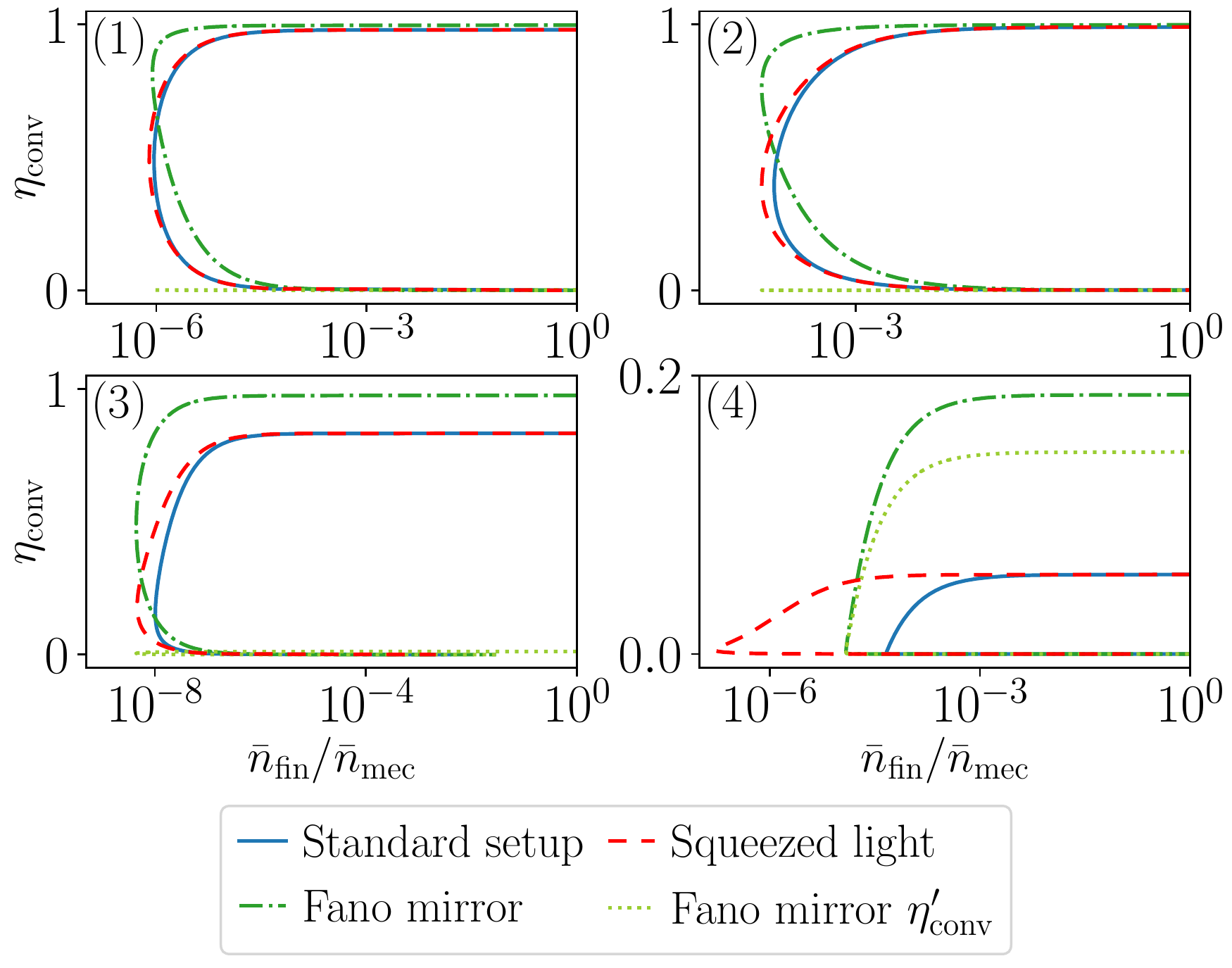}
 \caption{\label{fig:loop_etaC}
     Conversion efficiency $\etaC$ as a function of the effective phonon number for the three setups (indicated by different lines in each plot). The number in each sub-figure corresponds to the system number in parameter Tab.~\ref{tab:Params}. 
 }
\end{figure}
 
This is different only for system (4), which is not in the sideband-resolved regime. For system (4), as expected, the conversion efficiency is smaller than for the other systems. Surprisingly, the Fano-mirror setup is significantly more efficient than the other two setups, even when looking at $\etaC'$, which means that the asymmetry in the density of states of the cavity created by the Fano mirror can provide the same evacuated-heat flow out of the non-sideband resolved system, at the cost of much less resources than what is required for the standard setup using coherent or squeezed light.

The relation between the conversion efficiency and the smallest reachable steady-state phonon number, $\Neff$, is plotted in Fig.~\ref{fig:loop_etaC}. While also here a trade-off can be observed, importantly, the conversion efficiency is shown to be large down to phonon numbers that are very close to the lowest reachable values.

\section{Conclusion}\label{sec:IV}

In this work, we have provided a detailed thermodynamic analysis of optomechanical sideband cooling for different setups of cavity optomechanics---a standard setup employing a coherent laser drive or a squeezed laser drive, as well as a setup with a Fano mirror as one of the cavity mirrors. We have demonstrated how optomechanical cooling is realized by powering a heat valve that connects the mechanical resonator, heated by its coupling to a mechanical bath, to an engineered cold phonon bath provided by the optical driving. We have analyzed the lowest steady-state phonon number that can be reached for the mechanical system, together with the evacuated-heat flow and the efficiencies of the cooling process. The latter are additional, complementary performance quantifiers that we have identified in our heat-valve setting and that turn out to be particularly insightful in the analysis put forward here. 

This comprehensive theoretical analysis, which goes beyond the validity regime of previous studies of sideband cooling \cite{Asjad2016Nov}, employs realistic numbers from a broad range of possible experimental implementations \cite{thompson_strong_2008,groblacher_demonstration_2009,Eichenfield2009Nov, Chan2011,Delic2020,Delic2020_1,Rossi2018}. Our analysis has two main benefits. First, it provides an improved understanding of the underlying processes and the required mechanisms and resources for different optomechanical cooling setups. Second, the understanding of the \textit{thermodynamics} underlying sideband cooling, as discussed in this work, will enable further development of thermodynamic machines exploiting the heat flow between hot phonon bath and cold bath, engineered by the laser-light induced phonon-photon coupling. 

\section*{Acknowledgments}
We thank Alexia Auff\`eves, Andrew Jordan and Nikolai Kiesel for helpful discussions. We acknowledge funding from the Knut and Alice Wallenberg foundation through two Wallenberg Academy fellowships (J.S. and W.W.), from the Vetenskapsr\r{a}det, Swedish VR, under project numbers 2018-05061 (J.S. and J.M.) and 2019-04946 (W.W.) and the QuantERA project C’MON-QSENS! This work was performed in the context of the Excellence Initiative Nano at Chalmers.

\appendix

\section{Specific parameters used in the figures}

All the parameters used in the figures are gathered in Tab.~\ref{tab:Params}. The columns correspond to each of the studied systems. The uppermost section of the table contains the actual experimental parameters, the second one gives some insightful parameter ratios, the third one corresponds to the assumed parameters we used for the squeezed-light and Fano-mirror setups and the last one indicates the minimum achievable phonon number $\Neff$ in the mechanical resonator for each of the setups.

\begin{table*}[htb!]\setlength{\tabcolsep}{3pt}\renewcommand{\arraystretch}{1.6}
    \centering
    \begin{tabularx}{\linewidth}{>{\footnotesize}c c |X X X X }
        \hline
        \normalsize Quantity of interest&Symbol& System (1)  & System (2)  & System (3) & System (4)  \\     
        &&  &  \cite{Eichenfield2009Nov, Chan2011} &  \cite{Delic2020,Delic2020_1} &  \cite{Rossi2018} \\
        \hline\hline
          &  &\footnotesize sideband-resolved, MHz &\footnotesize sideband-resolved, GHz &\footnotesize sideband-resolved, levitated &\footnotesize non-sideband-resolved \\
        \hline\hline 
        Mechanical frequency &     $\Om/2\pi$ & $1\cdot10^6$\,Hz     &  $ 3.7\cdot 10^{9}$\, Hz  & $ 3.05\cdot 10^{5}$\, Hz & $ 1.14\cdot 10^{6}$\, Hz\\
        \hline
        Mechanical damping rate &     $\gamma/2\pi$   & 0.1\,Hz     &  $3.5\cdot 10^{4}$\,Hz   & $1.6\cdot 10^{-4}$ \,Hz & $1.1\cdot 10^{-3}\,$ Hz\\
        \hline
        Mechanical environment temperature&     $T_\m$  & 4\,K       &   $20$\,K     & $300$\,K      & $10$\,K\\
        \hline
        Average phonon number at $T_\m$ &     $\Nm$  & $8.3\cdot 10^4$      &   $7.1\cdot10^{2}$ &   $1.3\cdot10^{8}$ &   $1.1\cdot10^{6}$\\
        \hline
        Laser wavelength&     $\lambda_\text{las}$  & 1550\,nm       &   $1537$\,nm &   $1064$\,nm &   $795$\,nm\\
        \hline
        Cavity loss rate&     $\kappa/2\pi$    & $2\cdot10^5$\,Hz     &   $5\cdot10^{8}$\,Hz  &   $1.93\cdot10^{5}$\,Hz  &   $1.59\cdot10^{7}$\,Hz\\
        \hline
        Cavity length&     $L_\text{cav}$    & $3.75\cdot10^{-3}$\,m    &   $3\cdot10^{-6}\,$m  &  $1.07\cdot10^{-2}$\,m  &   $1.6\cdot10^{-3}$\,m\\
        \hline
        Cavity free spectral range &     $\Gamma/2\pi$    &  $4.0\cdot10^{10}$\,Hz &  $1.44\cdot10^{13}$\,Hz    &  $1.40\cdot10^{10}$\,Hz  &   $9.38\cdot10^{10}$\,Hz\\
        \hline
        Single-photon coupling strength &     $g_0/2\pi$  & 10\,Hz      &   $9.1\cdot10^{5}$\,Hz &   $0.3$\,Hz &   $129$\,Hz\\
        \hline
        \hline
        Sideband resolution &     $\Om/\kappa$ & $5$      &   $7.4$ &   $1.58$ &   $0.07$\\
        \hline
        Granularity parameter&     $g_0/\kappa$ &   $5.0\cdot10^{-5}$    & $ 1.8\cdot 10^{-3}$ & $ 1.5\cdot 10^{-6}$ & $ 8.1\cdot 10^{-6}$\\
        \hline
        Mechanical quality factor &     $Q_\m=\Om/\gamma$ & $1.0 \cdot 10^7$      & $ 1\cdot 10^{5}$ & $ 1.9\cdot 10^{9}$ & $ 1.03\cdot 10^{9}$\\
        \hline
        Single-photon quantum cooperativity &     $C_0 = 2g_0^2/\kappa\gamma\Nm$   &  $ 6.0\cdot 10^{-8}$  & $ 6.6\cdot 10^{-5}$ & $ 2.2\cdot 10^{-11}$ & $ 8.6\cdot 10^{-7}$\\
        \hline
        Number of coherent oscillations &     $Q_\m/\Nm$   & $120.0$  & $ 140.8$ & $ 14.8$ & $ 936.7$\\
        \hline\hline
        Effective detuning  &    $\Delta=\Delta_0-g_0 \sqrt{2}\bar{q}$ & $\Om$      & $\Om$  & $\Om$  & $ \kappa$\\
        \hline
        Squeezing angle  &    $\theta$ & $0.835$ rad     & $0.819$ rad  & $0.939$ rad  & $1.11$ rad\\
        \hline
        Squeezing ratio  &    $r_\s$ & $0.050$     & $0.034$  & $0.154$  & $0.711$\\
        \hline
        Squeezing level\footnote{This is the amount of squeezing in dB below shot noise, with the variance of shot noise being $1/2$ and $\mean{\Delta P^\theta_\text{in,s}}$ the variance of the squeezed quadrature.}  & $\!\!\!-10\log_{10}(2\mean{\Delta P^\theta_\text{in,s}})$ & 0.87\,dB & 0.59\,dB & 2.7\,dB & 15.4\,dB\\
        \hline
        Fano mode loss rate&     $\gamma_d/2\pi$    & $8.0\cdot10^7$\,Hz     &   $4.38\cdot10^{11}$\,Hz  &   $7.71\cdot10^{6}$\,Hz  &   $1.31\cdot10^{6}$\,Hz\\
        \hline
        Fano setup, left-side loss rate &     $\kappa_\L/2\pi$    & $8.0\cdot10^{10}$\,Hz     &   $2.88\cdot10^{13}$\,Hz  &   $2.80\cdot10^{10}$\,Hz  &   $1.88\cdot10^{11}$\,Hz\\
        \hline
        Fano setup, right-side loss rate &     $\kappa_\R/2\pi$    & $5.0\cdot10^7$\,Hz     &   $8.22\cdot10^{9}$\,Hz  &   $1.75\cdot10^{8}$\,Hz  &   $5.70\cdot10^{11}$\,Hz\\
        \hline\hline
        Minimum phonon number, standard setup & $\min(\Neff^\text{standard})$ & 0.077& 0.021& 0.21 & 6.7 \\
        \hline
        Minimum phonon number, squeezed-light setup & $\min(\Neff^\text{suqeezing})$& 0.066& 0.016& 0.096& 0.039 \\
        \hline
        Minimum phonon number, Fano-mirror setup &$\min(\Neff^\text{Fano})$& 0.073& 0.016& 0.091& 2.4 \\
        \hline
    \end{tabularx}
    \caption{\label{tab:Params}
        Parameters of four different optomechanical systems that we consider for our thermodynamic analysis of optomechanical cooling. The last section in the table indicates the minimum achievable steady-state phonon number $\Neff$ for each of the setups and the section above corresponds to the assumed parameters we used in the figures. 
    }
\end{table*}

\section{Standard setup} \label{appendix:standard}

\subsection{Steady state and linearization} \label{appendix:standard:lin}
The semi-classical steady state of the optomechanical system is given by Eqs.~\eqref{standard:average values}, which are actually a non-linear system of equations because the effective detuning $\Delta$ depends on the steady-state mechanical position $\bar{q}$~\cite{Genes2008Mar}. As a consequence, the average photon number in the cavity, $\Nc$,  is a solution of the cubic equation
\begin{equation}
    \Nc\left(\kappa^2 + \left(\Delta_0 - \frac{2g_0^2}{\Om}\Nc\right)^2\right) = \abs{\epsilon}^2.
\end{equation}
Therefore there are two different regimes depending on the optomechanical parameters and the laser power and detuning: one where the above equation has a single real root and one where it has three. The systems (see Tab.~\ref{tab:Params})  with the respective detunings considered in this article are in the first regime, therefore there is a unique steady state $(\alpha, \bar{q})$. On the other hand, in the second regime there are multiple possible steady states, see for instance~\cite{Bowen2015Nov}, Chap.~2, for more details.

We study the evolution of the optomechanical system around this steady state by splitting the Heisenberg operators in the Langevin equations~\eqref{standard:Langevin} into a steady-state value and a fluctuation operator: $\hat{a} = \alpha + \da$, $\hat{q} = \bar{q} + \dq$ and $\hat{p} = \bar{p} + \deltp$, with $\bar{p} = 0$. Then, we use Eqs.~\eqref{standard:average values} and neglect the second order terms (in $\da\dq$ and $\da^\dagger \da$) to obtain the linearized Langevin equations~\eqref{standard:Langevin lin}. In these equations, the non-dissipative part of the evolution is governed by the quadratic Hamiltonian $\hat{H}_\text{lin}$, given in Eq.~\eqref{H_Lin}.

\begin{figure}[htb]
    \includegraphics[width=\linewidth]{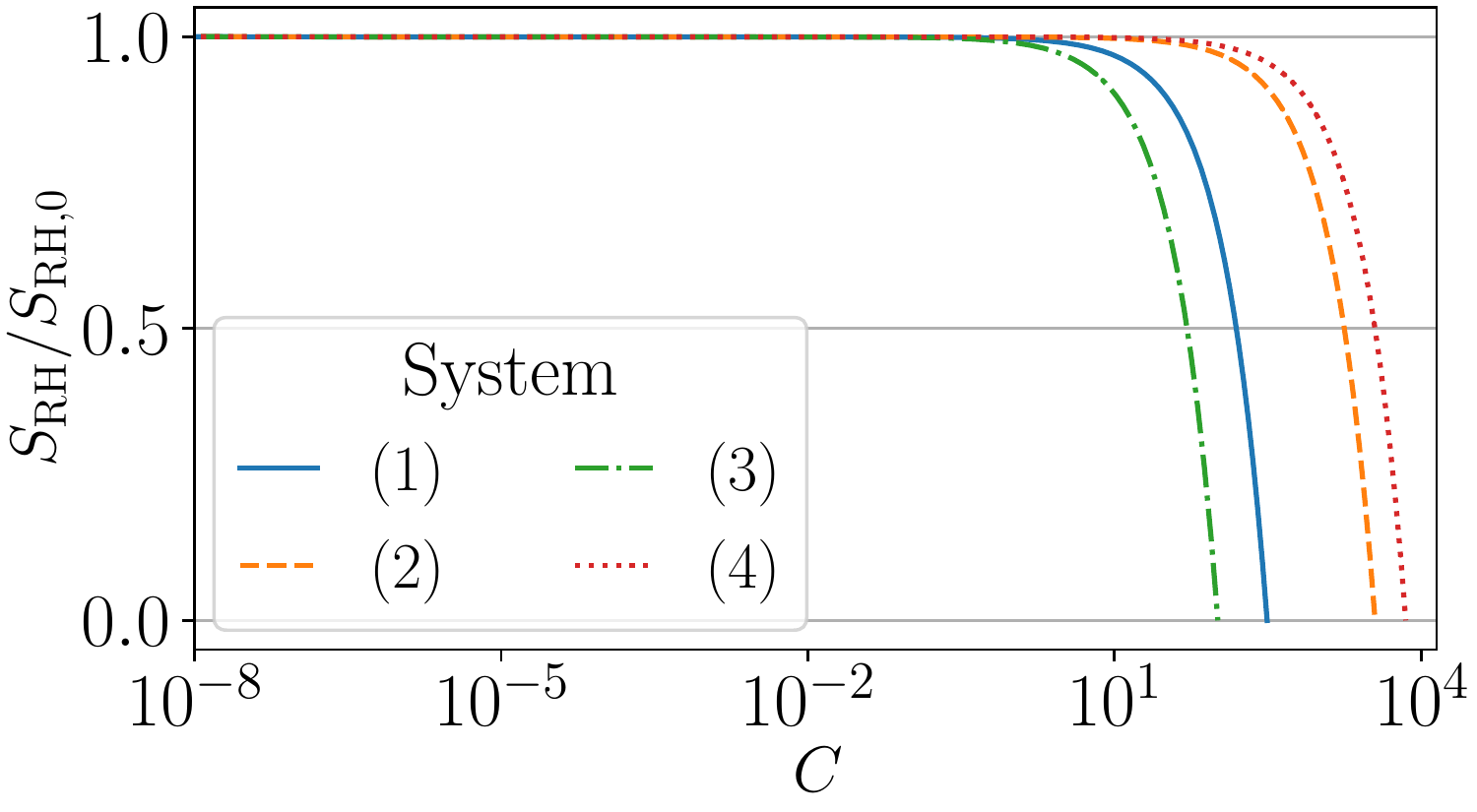}
    \caption{\label{fig:appendix stability}
         Stability condition $S_\text{RH}$ as a function of the cooperativities considered in this article for each of the systems from Tab.~\ref{tab:Params}. $S_\text{RH}$ has been normalized by its value at zero coupling, $S_{\text{RH}, 0} = \Om(\kappa^2 + \Delta^2)$.
    }
\end{figure}
The stability of the optomechanical system can be determined by applying the Routh-Hurwitz criterion, from which emerges the stability condition $S_\text{RH} > 0$ \cite{Wilson-Rae2008Sep, Genes2008Mar}, with
\begin{equation}
    S_\text{RH} = \Om(\kappa^2 + \Delta^2) - 4g^2\Delta. \label{standard:stability criterion}
\end{equation}
As can be seen in Fig.~\ref{fig:appendix stability}, the four systems studied in this article fulfill this stability condition for all the considered values of the coupling $g$.

\subsection{Solving the dynamics}\label{appendix:standard:Langevin}

\subsubsection{Solution of the Langevin equations} \label{appendix:standard:solution Langevin}
The system of equations~\eqref{standard:Langevin lin} can be solved in the frequency domain~\cite{Genes2008Mar}. We have used the following convention for the Fourier transform: $\hat{f}[\omega] = \int_{-\infty}^{+\infty} \dd t \e^{i\omega t}\hat{f}(t)$. In particular, the expression of the mechanical position is given by
\begin{align} \label{standard:solution Langevin}
    \chi_\m^\eff[\omega]^{-1}\dq[\omega] = & \frac{\sqrt{2\kappa} 2g\left((\kappa - i\omega)\hat{X}_\text{in}[\omega] + \Delta\hat{P}_\text{in}[\omega] \right)}{(\kappa - i\omega)^2 + \Delta^2}\nonumber \\
    &+ \sqrt{\gamma}\hat{\xi}[\omega], 
\end{align}
with the input noise quadratures
 $\hat{X}_\text{in} = (\ain + \ain^\dagger)/\sqrt{2}$ and $\hat{P}_\text{in} = i(\ain^\dagger - \ain) /\sqrt{2}$.
The mechanical susceptibility is given by
\begin{equation}\label{standard:susceptibility}
    \chi_\m^\eff[\omega] = \Om\left(\Om^2 - \omega^2 - i\omega\gamma -\frac{4g^2\Delta\Om}{(\kappa -i\omega)^2 + \Delta^2}\right)^{-1}\!\!\!.
\end{equation}
We can put it in the usual form for a harmonic oscillator~\cite{Aspelmeyer2014Dec}, that is $\chi_\m^\eff[\omega] = \Om\left(\Om^\eff[\omega]^2 - \omega^2 - i\omega\gamma^\eff[\omega]\right)^{-1}$, identifying the effective mechanical frequency $\Om^\eff[\omega] = \Om + \delta\Om[\omega]$ and damping rate $\gamma^\eff[\omega] = \gamma + \Gopt[\omega]$. We have denoted
\begin{align}
    \delta\Om[\omega] &= - \frac{2g^2\Delta(\kappa^2 - \omega^2 + \Delta^2)}{(\kappa^2 + (\omega - \Delta)^2)(\kappa^2 + (\omega + \Delta)^2)},
    \\
    \Gopt[\omega] &= \frac{8g^2\Delta\Om\kappa}{(\kappa^2 + (\omega - \Delta)^2)(\kappa^2 + (\omega + \Delta)^2)}.\label{standard:Gopt}
\end{align}
$\delta\Om[\omega]$ is the frequency shift due to the optical spring effect~\cite{Aspelmeyer2014Dec}.
In the weak-coupling regime, we can neglect the contributions to the evolution of the non-resonant frequencies, therefore the effective frequency from the master equation~\eqref{weak-coupling:master_eq} is equal to $\Om^\eff[\Om]$ and the cold bath damping rate is given by $\Gopt[\Om]$.

\subsubsection{Lyapunov equation}\label{appendix:standard:Lyapunov}
From the Langevin equations~\eqref{standard:Langevin lin} and using the correlation functions of the noise, Eqs.~\eqref{correlations_xi} and \eqref{standard:correlations_a_in}, we obtain the evolution of the second-order moments of the quadratures:
{\allowdisplaybreaks
\begin{widetext}
    \begin{subequations}\label{standard:eqs 2nd order moments}
        \begin{align}
            \dv{\mean{\dXa^2}}{t} =\,& 2 {\Delta} \mean{\dXa\dPa} - 2 \kappa \mean{\dXa^2} + \kappa, \\
            \dv{\mean{\dXa\dPa}}{t} =\,& {\Delta} (\mean{\dPa^2} - \mean{\dXa^2}) + 2 g \mean{\dXa\dq} - 2 \kappa \mean{\dXa\dPa},  \\
            \dv{\mean{\dXa\dq}}{t} =\,& \Om \mean{\dXa\deltp} + {\Delta} \mean{\dPa\dq} - \kappa\mean{\dXa\dq},  \\
            \dv{\mean{\dXa\deltp}}{t} =\,& -\Om \mean{\dXa\dq} + {\Delta} \mean{\dPa\deltp} + 2 g \mean{\dXa^2}  - (\gamma + \kappa) \mean{\dXa\deltp}, \\
            \dv{\mean{\dPa^2}}{t} =\,& -2 {\Delta} \mean{\dXa\dPa} + 4 g \mean{\dPa\dq} - 2 \kappa \mean{\dPa^2} + \kappa, \\
            \dv{\mean{\dPa\dq}}{t} =\,& -{\Delta} \mean{\dXa\dq} + \Om \mean{\dPa\deltp} - \kappa\mean{\dPa\dq}  + 2 g \mean{\dq^2}, \\
            \dv{\mean{\dPa\deltp}}{t} =\,& -{\Delta} \mean{\dXa\deltp} - \Om \mean{\dPa\dq}  - (\gamma + \kappa)\mean{\dPa\deltp} + 2 g( \mean{\dXa\dPa} +  \mean{\dq\deltp}), \\
            \dv{\mean{\dq^2}}{t} =\,& 2  \Om \mean{\dq\deltp}, \\
            \dv{\mean{\dq\deltp}}{t} =\,& \Om (\mean{\deltp^2} - \mean{\dq^2}) + 2 g \mean{\dXa\dq} - \gamma\mean{\dq\deltp}, \\
            \dv{\mean{\deltp^2}}{t} =\,& -2  \Om \mean{\dq\deltp} + 4 g \mean{\dXa\deltp}   - 2 \gamma \mean{\deltp^2} + {\gamma}(2 \Nm + 1),
        \end{align}
    \end{subequations}
\end{widetext}}
\noindent where we have defined the optical quadratures $\dXa = (\da + \da^\dagger)/\sqrt{2}$ and $\dPa = (\da - \da^\dagger)/i\sqrt{2}$.
From these equations, we get the evolution of the photon and phonon numbers, $\mean{\da^\dagger \da} = \frac{1}{2}(\mean{\dXa^2} + \mean{\dPa^2} - 1)$ and $\mean{\db^\dagger \db} = \frac{1}{2}(\mean{\dq^2} + \mean{\deltp^2} - 1)$:
\begin{subequations}\label{2nd_order_moments_meq}
    \begin{align}
        \dv{\mean{\da^\dagger \da}}{t} &= 2 g \mean{\dPa\dq} - 2\kappa\mean{\da^\dagger \da},\\ 
        \dv{ \mean{\db^\dagger \db}}{t} &= 2 g \mean{\dXa\deltp} + \gamma \left(\Nm + \frac{1}{2}- \mean{\deltp^2} \right)\nonumber\\
        &=-I^\text{phonon}_\text{c} -I^\text{phonon}_\text{h}.
    \end{align}
\end{subequations}
We have identified the phonon flows $I^\text{phonon}_\text{c} = - 2 g \mean{\dXa\deltp}$ and $I^\text{phonon}_\text{h} = - \gamma \left(\Nm + \frac{1}{2}- \mean{\deltp^2} \right)$ directed towards the cold and hot bath, respectively. Similarly, the flow of photons coming from the cavity and interacting with the mechanics reads $I^\text{photon} = 2 g \mean{\dPa\dq}$.
Rewriting Eqs.~\eqref{2nd_order_moments_meq} fully in terms of the annihilation and creation operators, we obtain Eqs.~\eqref{standard:evolution photon and phonon numbers}.

The linearized Hamiltonian $\hat{H}_\text{lin}$ (Eq.~\eqref{H_Lin}) is Gaussian. Therefore, the system of equations~\eqref{standard:eqs 2nd order moments} can be rewritten in the form of a Lyapunov equation for $V$, the covariance matrix of the quadratures,
    \begin{align}\label{standard:Lyapunov equation}
        \dv{V}{t} = A V+V A^T+ B.
    \end{align} 
The elements of covariance matrix $V$ are defined as 
\begin{align}\label{cov matrix elements}
    V_{ij}= \dfrac{1}{2} \mean{\{ Y_i, Y_j\}}  - \mean{Y_i} \mean{Y_j},
\end{align}
where $\vec{Y}$ is the vector of operators $\left( \dXa, \dPa, \dq, \deltp  \right)$.
We have defined the matrices\\
\begin{subequations}
    \begin{align}
    A&= \begin{bmatrix}
        - \kappa  & \Delta & 0 & 0  \\
        -\Delta&  - \kappa  & 2g & 0  \\ 
        0 & 0 & 0 & \Om \\
        2g & 0 & -\Om&  -\gamma
    \end{bmatrix},\label{standard:Lyapunov:A}\\
    B&= \begin{bmatrix}
        \kappa & 0 & 0 & 0  \\ 
        0& \kappa  & 0 & 0  \\ 
        0& 0 & 0 & 0 \\
        0 & 0 & 0& \gamma \left(2\bar{n}+1\right) 
    \end{bmatrix}\label{standard:Lyapunov:B}.
\end{align}
\end{subequations}
We then determine the steady-state $\bar{V}$ by solving $0 = A \bar{V}+\bar{V} A^T+ B$ and obtain the steady-state phonon number
\begin{equation}
    \Neff = \frac{1}{2}(\bar{V}_{33} + \bar{V}_{44} - 1). \label{neff_Lyapunov}
\end{equation}
Furthermore, the steady-state efficiencies (Eqs.~\eqref{etaC} and \eqref{etaL}) can be expressed as
\begin{align}
    \etaL &= -\frac{4\kappa\Om g\bar{V}_{14}}{\omL \abs{\epsilon}^2},\label{etaL_Lyapunov}\\
    \etaC &=  -\frac{\bar{V}_{14}}{\bar{V}_{23}}\label{etaC_Lyapunov}.
\end{align}
We therefore obtain the following analytical expressions
\begin{align}
\Neff &= \frac{g^2 c_1 + \gamma\Nm c_2}{c_3  S_\text{RH}},
\\
\etaL &= \frac{(4\Om g_{0}^{2} {\gamma} {\kappa}^{2})(c_4 + \Nm c_5)}{{\left({\Delta}^{2} + {\kappa}^{2}\right)} \omL c_3},\\
 \etaC &= \frac{\gamma  S_\text{RH} (c_4 + \Nm c_5)}{\Om(c_6 + \gamma\Nm c_7)},
\end{align}
with  $S_\text{RH}$ defined by Eq.~\eqref{standard:stability criterion} and
{\allowdisplaybreaks
\begin{widetext}
    \begin{subequations}\label{standard:full analytical expr}
    \begin{align}
        c_1 =\,
         &\Big[ {\Om}  ({\Delta}^{2} + {\kappa}^{2})^2 + {\left({\Delta}^{2} - 4 \, {\Delta} {\Om} + 2 \, {\Om}^{2} + {\kappa}^{2}\right)} S_\text{RH} \Big] \kappa^2  \nonumber\\
        &+ \gamma \Big[{\Delta}^{2}({\Delta}^{3} + {\Om}^{3}) - 2 \, {\Delta} {\Om}^{2} g^{2} + (2 {\Delta}^{3} + 2 {\Delta}^{2} {\Om} + {\Om}^{3}) {\kappa}^{2}+ {\left({\Delta} + 2 \, {\Om}\right)} {\kappa}^{4}\nonumber\\
        &\hspace{0.8cm} + {\left({\Delta}^{2} - 3 \, {\Delta} {\Om} + {\kappa}^{2}\right)} S_\text{RH} \Big]\kappa\nonumber\\
        &+ \gamma^2\Big[ 2  {\Delta}^{2} \Om g^{2} + {\left(2  {\Delta} + \Om\right)} {\kappa}^{4} +  {\Delta}^{2}{\left(2  {\Delta} + \Om\right)} {\kappa}^{2}\Big] + \gamma^3 (\Delta^2 + \kappa^2)\kappa\Delta,\\
        c_2 =\,
        & \Big[ \Om {\kappa}^{6} + {\left(3 {\Delta}^{2} \Om + 2 \Om^{3} - 2 {\Delta} g^{2}\right)} {\kappa}^{4} + {\left((3 {\Delta}^{4}  + \Om^{4})\Om - 2(2 {\Delta}^{2}  +3  \Om^{2}) \Delta g^{2}\right)} {\kappa}^{2} \nonumber\\
        &\hspace{0.2cm}+ \Delta{\left(({\Delta}^{2} -  \Om^{2})^2\Delta\Om - 2({\Delta}^{4}- 5 {\Delta}^{2} \Om^{2}   +2  \Om^{4} ) g^{2} - 8 {\Delta} \Om g^{4}\right)}\Big] \kappa\nonumber\\
        & + \gamma\Big[ 2 \Om {\kappa}^{6} + 2 {\left(2 {\Delta}^{2} \Om + \Om^{3} - 2 {\Delta} g^{2}\right)} {\kappa}^{4}+ 2\Delta {\left(\Delta\Om ({\Delta}^{2} + \Om^{2}) -( 2 {\Delta}^{2} + 3 \Om^{2} )g^{2}\right)} {\kappa}^{2} \nonumber\\
        &\hspace{0.75cm} + 2 g^2 {\Delta}^{2} \Om \left({\Delta} \Om - 2  g^{2} \right)\Big]  +\gamma^2 \Big[\Om {\kappa}^{4} + 2 {\left({\Delta} \Om -  g^{2}\right)}\Delta {\kappa}^{2} + {\left({\Delta} \Om - 2  g^{2}\right){\Delta}^{3}}\Big]\kappa,\\
        c_3 =\,
        &8 {\Delta} \Om g^{2} {\kappa}^{2}   + \gamma \left[{\kappa}^{4} + 2 {\left({\Delta}^{2} + \Om^{2}\right)} {\kappa}^{2} + {({\Delta}^{2} - \Om^{2})^2  + 8 {\Delta} \Om g^{2}} \right]\kappa \nonumber\\
        & + 2{\gamma}^{2}  {\left[{\Delta} \Om g^{2} + ({\Delta}^{2} + \Om^{2}) {\kappa}^{2} + {\kappa}^{4}\right]} + {\gamma}^{3}  {\left({\Delta}^{2}  + {\kappa}^{2}\right)} \kappa\\
        c_4 =\,
        & - \left({\kappa}^{2} + {\left({\Delta} - \Om\right)}^2 \right){\kappa} + {\left( {\Delta} \Om + {\kappa}^{2}- {\Delta}^{2}  \right)} {\gamma} \\
        c_5 =\,
        & 2  {\Delta} \Om {\left({\gamma} + 2  {\kappa}\right)}\\
        c_6 =\,
        & 4{\left({\Delta}^{2} + {\kappa}^{2}\right)} g^{2} {\kappa}^{2} + \gamma \Big[  ({\kappa}^{2} + \Delta^2)^2 +2g^2(4\kappa^2 + \Om^2) + (\Om - 2\Delta)S_\text{RH} \Big]\kappa \nonumber \\
        & + \gamma^2 \Big[ 2{\kappa}^{4} + {\left({\Delta}^{2} + ({\Delta} - \Om)^{2} + 4g^{2}\right)} {\kappa}^{2} + {\Delta}^{2} \Om(\Om-{\Delta})+ 2 {\Delta} {\left(2{\Delta} -  \Om\right)} g^{2} \Big]  + \gamma^3 {\left({\Delta}^{2} + {\kappa}^{2}\right)} {\kappa} \\
        c_7 =\,
        &  \Big[ ({\kappa}^{2} + \Delta^2)^2 + \Om S_\text{RH} \Big]2\kappa 
         + \gamma \Big[ (\Om^2 + 4\kappa^2)(\kappa^2 + \Delta^2) + \Om S_\text{RH}\Big] + \gamma^2 {\left({\Delta}^{2} + {\kappa}^{2}\right)} 2 {\kappa}\\\nonumber
    \end{align}
    \end{subequations}
\end{widetext}}

When $g$ reaches the critical value 
\begin{equation}
    g_\textbf{crit} = \sqrt{\frac{\Om}{4\Delta}(\kappa^2 + \Delta^2)},
\end{equation}
the matrix $A$ is not invertible and the Lyapunov equation \eqref{standard:Lyapunov equation} does not have a steady state. In particular, $\Neff$ diverges, as visible in Fig.~\ref{fig:neff}.  This critical value of the coupling $g$ corresponds to the point at which the stability condition $S_\text{RH} > 0$ (see Appendix \ref{appendix:standard:lin}) is no longer fulfilled. 

\subsubsection{Beyond the white-noise approximation}\label{appendix:standard:beyond white noise}
   According to~\cite{Giovannetti2001Jan}, the white-noise approximation for the mechanical noise (Eq.~\eqref{correlations_xi}) gives inconsistent results, even at high temperature because the commutation relation $[\hat q, \hat p] = i$ is not preserved and it yields a spurious term in the phase fluctuation spectrum. Depending on the quantities of interest this does not necessarily impact the results, like in our case (see Appendix \ref{appendix:standard:approxs}). Nevertheless, to restore the correct commutation relation, the thermal noise operator should fulfill the following relations
   \begin{align}
     \!\! \!\mean{\hat\xi(t)\hat\xi(t')} &= \int\!\!\dd \omega \frac{\omega\e^{- i \omega(t-t')}}{2\pi} \frac{\coth(\frac{\hbar\omega}{2\kB T_\m})+1}{\Om},\!\!\\
      \!\!\! [\hat\xi(t), \hat\xi(t')] &= 2i\frac{1}{\Om}\dv{t}\delta(t-t').
   \end{align}
   To compute the steady-state phonon number without making the white-noise approximation, we use the same method as in~\cite{Genes2008Mar}. 
   The effective phonon number is given by
       $\frac{1}{2}(\mean{\dq^2} + \mean{\deltp^2} - 1)$,
   therefore we can compute it from the position and momentum fluctuations:
   \begin{align}
       \mean{\dq^2} &= \int_{-\infty}^{+\infty}\frac{\dd \omega}{2\pi} S_q[\omega],\\
       \mean{\deltp^2} &= \int_{-\infty}^{+\infty}\frac{\dd \omega}{2\pi}\frac{\omega^2}{\Om^2} S_q[\omega].
   \end{align}
   The position fluctuation spectrum,
   \begin{align}
       S_q[\omega] &=\int_{-\infty}^{+\infty} \dd t \e^{i\omega t}\mean{\dq(t)\dq(0)}\nonumber\\
       &= \int_{-\infty}^{+\infty}\frac{\dd\omega'}{2\pi}\mean{\dq[\omega]\dq[\omega']}, \label{fluctuation spectrum}
   \end{align}
   is obtained from the solution of the Langevin equations~\eqref{standard:solution Langevin}.
   In this case, the noise correlation functions are
   \begin{equation}
       \mean{\hat{\xi}[\omega]\hat{\xi}[\omega']} = 2\pi\frac{\omega}{\Om}\left[ 1 + \coth(\frac{\hbar\omega}{2\kB T_\m})\right] \delta(\omega + \omega'),
   \end{equation}
   and
   \begin{subequations}\label{correlation cavity input noise}
       \begin{align}
       \hspace*{-0.3cm}\mean{\hat{X}_\text{in}[\omega]\hat{X}_\text{in}[\omega']} &=  \mean{\hat{P}_\text{in}[\omega]\hat{P}_\text{in}[\omega']} = \pi\delta(\omega + \omega')\!\!\\
       \hspace*{-0.3cm}\mean{\hat{X}_\text{in}[\omega]\hat{P}_\text{in}[\omega']} &= -\mean{\hat{P}_\text{in}[\omega]\hat{X}_\text{in}[\omega']}  =  i\pi\delta(\omega + \omega')\!\!\!
   \end{align}
   \end{subequations}
   So, we get 
   \begin{equation}
       S_q[\omega] = \abs{\chi_\m^\eff[\omega] }^2(S_\text{th}[\omega] + S_\text{rp}[\omega])\label{Sq}
   \end{equation}
   where
   \begin{align}
       S_\text{th}[\omega] &= \gamma\frac{\omega}{\Om}\coth(\frac{\hbar\omega}{2\kB T_\m}), \label{Sth}\\
       S_\text{rp}[\omega] &= \frac{4g^2\kappa}{\kappa^2 + (\omega - \Delta)^2},\label{Srp}
   \end{align}
   are the thermal and radiation pressure noise spectra. To avoid lengthy computation we did not calculate the analytical expression for the phonon number, but integrated it numerically instead. We confirmed that it gives almost identical results as the white-noise approximation for the parameters considered in this article [see Fig.~\ref{fig:appendix approx comparison}].\\
   
   For the other quantities, namely $J_\text{c}$, $\etaL$ and $\etaC$, we would also need to compute $\mean{\dXa\deltp}$ and $\mean{\dPa\dq}$. This can be done by inserting the solution of the Langevin equations~\eqref{standard:Langevin lin} in the frequency domain into 
   \begin{align}
       \mean{\dXa\deltp} &= \int_{-\infty}^{+\infty}\frac{\dd \omega}{2\pi}\int_{-\infty}^{+\infty}\frac{\dd \omega'}{2\pi}\mean{\dXa[\omega]\deltp[\omega']},\\
       \mean{\dPa\dq} &= \int_{-\infty}^{+\infty}\frac{\dd \omega}{2\pi}\int_{-\infty}^{+\infty}\frac{\dd \omega'}{2\pi}\mean{\dPa[\omega]\dq[\omega']}.
   \end{align}      
   Note that this approach could be used to treat a case where the input noise in the cavity is not vacuum noise by replacing Eqs.~\eqref{correlation cavity input noise} by the appropriate correlation functions. 

\subsection{Contributions to the evacuated-heat flow}\label{appendix:standard:heat flows}
\begin{figure}[htb]
    \includegraphics[width=\linewidth]{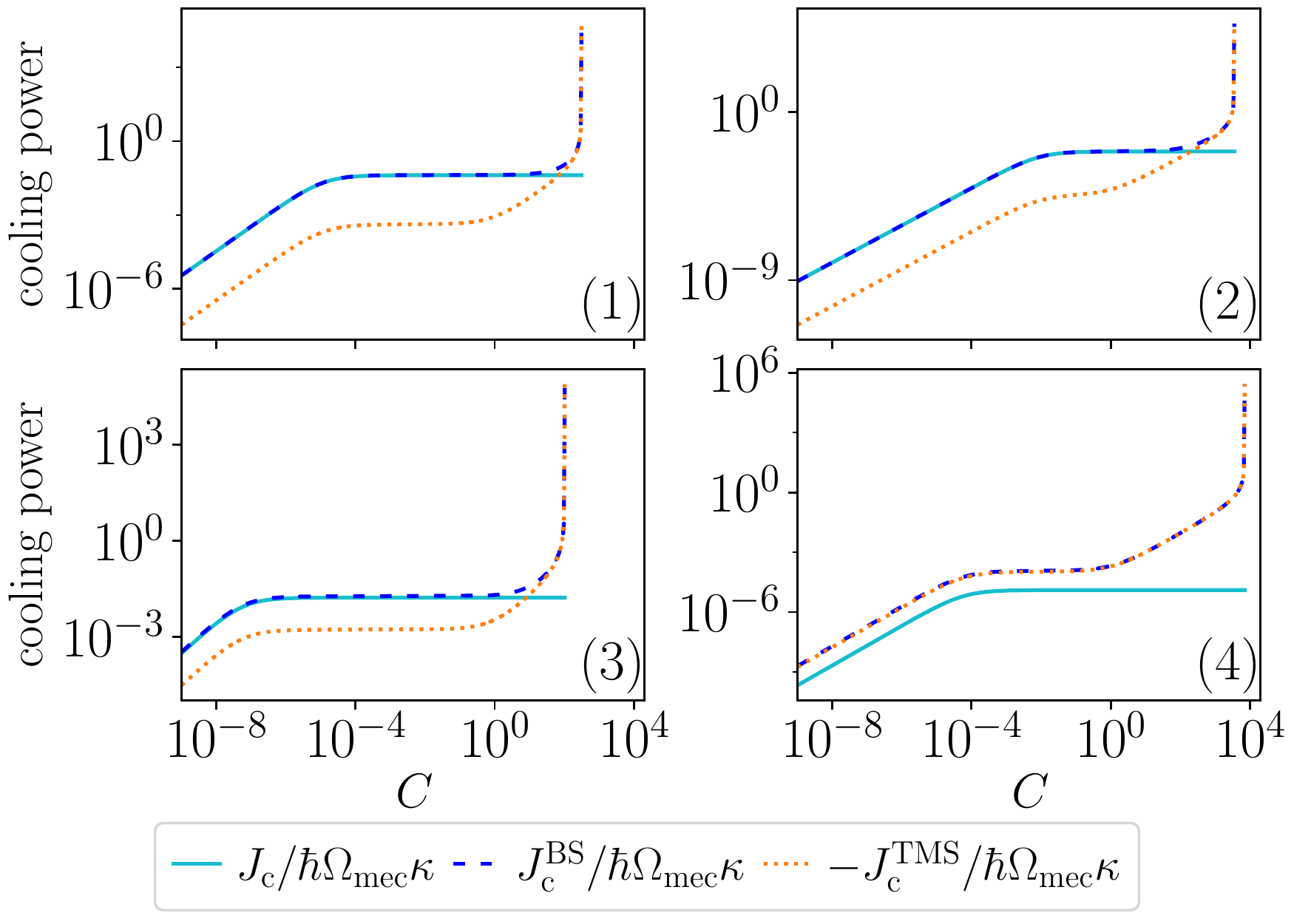}
    \caption{\label{fig: appendix Jc contributions}
        Beam splitter (dashed dark blue line) and two-mode squeezing (dotted orange line) contributions to the evacuated-heat flow $J_\text{c}$ (solid cyan line) as a function of cooperativity for each of the systems from Tab.~\ref{tab:Params}. We have actually plotted $-J_\text{c}^\text{TMS}$ so as to see all the curves on the same plot in log-log scale.
    }
\end{figure}

The evacuated-heat flow  $J_\text{c}$, Eq.~\eqref{cooling_power}, can be split into two contributions: $J_\text{c} = J_\text{c}^\text{BS} + J_\text{c}^\text{TMS}$, where $J_\text{c}^\text{BS} = i g\mean{\da^\dagger \db - \da \db^\dagger}$ comes from the beam-splitter interaction term $-\hbar g (\da^\dagger \db + \da \db^\dagger)$ in the Hamiltonian $\hat{H}_\text{lin}$ (Eq.~\eqref{H_Lin}) and $J_\text{c}^\text{TMS} = i g\mean{\da \db - \da^\dagger \db^\dagger}$ from the two-mode-squeezing term $-\hbar g (\da \db + \da^\dagger \db^\dagger)$. In the rotating wave approximation, valid in the weak coupling regime (see next section~\ref{appendix:standard:approxs}) for $\Delta$ close to $\Om$, only the beam-splitter contribution is taken into account. Figure~\ref{fig: appendix Jc contributions} indeed shows that in the resolved-sideband regime (systems (1) to (3)), $J_\text{c}^\text{TMS}$ is negligible at low $C$, then both contributions diverge while $J_\text{c}$ remains constant. In the non-resolved-sideband regime (system (4)), $J_\text{c}$ has the same behavior but the two contributions are always of the same order of magnitude.

\subsection{Measuring the photon and phonon flows}\label{appendix:standard:flows measurement}

In an experiment, the power spectra of the quadratures $\hat{X}_\text{out}$ and $\hat{P}_\text{out}$ of the light field leaking out of the cavity can be measured using a double homodyne detection scheme. They are related to the fluctuation spectra of the quadratures of the intra-cavity field by
\begin{align}
    S_{{X}_\text{out}}[\omega] &= S_{{X}_\text{in}}[\omega] + 4\kappa |\alpha|^2\delta(\omega)  + 2\kappa S_{{X}_a}[\omega],\\
    S_{{P}_\text{out}}[\omega] &= S_{{P}_\text{in}}[\omega] + 2\kappa S_{{P}_a}[\omega],
\end{align}
where we have used the input-output relation $\hat{a}_\text{out} = \ain - \sqrt{2\kappa}\hat{a}$. The power spectra are defined like in Eq.~\eqref{fluctuation spectrum}.

Besides, using the equations for the second-order moment  \eqref{standard:eqs 2nd order moments} in the steady-state, the phonon and photon flows, $I_\text{c}^\text{phonon}$ from Eq.~\eqref{cooling_power} and $I^\text{photon}$ from Eq.~\eqref{photon flow}, can be expressed as functions of $\mean{\dXa^2}$ and $\mean{\dPa^2}$:
\begin{align}
   \!\! I_\text{c}^\text{phonon} 
    &= \frac{2 \kappa }{\Delta\Omega}\left[ \kappa^2\mean{\dXa^2} - \Delta^2\mean{\dPa^2} +\frac{\Delta^2 - \kappa^2}{2}\right],\!\\
   \!\! I^\text{photon} 
    &=  \kappa \left[\mean{\dXa^2} + \mean{\dPa^2} - 1\right].
\end{align}
For a given operator $\hat{A}$, $\mean{\hat{A}^2} = \int_{-\infty}^{+\infty} \frac{\dd{\omega}}{2\pi} S_A[\omega]$,
so we obtain
\begin{align}
    I_\text{c}^\text{phonon} 
    =\,& \frac{ 2\kappa }{\Delta\Omega}\int_{-\infty}^{+\infty} \frac{\dd{\omega}}{2\pi} \left(\kappa^2 S_{{X}_a}[\omega]-\Delta^2 S_{{P}_a}[\omega]\right) \nonumber\\
    & + \frac{ \kappa (\Delta^2-\kappa^2)}{\Delta\Omega},\\
    I^\text{photon} 
    =\,&\kappa\int_{-\infty}^{+\infty} \frac{\dd{\omega}}{2\pi} \left( S_{{X}_a}[\omega] +  S_{{P}_a}[\omega]\right) - \kappa.
\end{align}
The cavity loss rate $\kappa$ and the effective detuning $\Delta$ can be measured using optomechanically induced transparency, therefore 
$I_\text{c}^\text{phonon}$ and $I^\text{photon}$ are accessible experimentally. Finally, since the laser input power $\Pin$ can also be determined, we  have access to the evacuated-heat flow $J_\text{c}$ and both efficiencies $\etaL$ and $\etaC$.

\subsection{Comparison of the different approximations}\label{appendix:standard:approxs}
Different approaches involving various approximations are used in the literature when studying optomechanical systems. 
Here we provide a quick overview of their differences and how this impacts the steady-state phonon number in the case of the setups considered in this article.

The simplest case is to consider the weak-coupling regime (see Ref.~\cite{Marquardt2007Aug, Aspelmeyer2014Dec} and Sec.~\ref{sec:II:Thermodynamic picture}) where $g \ll \kappa, \Om$. This allows to neglect the mechanical backaction in the fluctuation spectrum of the radiation pressure force and the steady-state phonon number in the resonator is given by Eq.~\eqref{weak-coupling:neff}.

When going beyond the weak coupling limit, two commonly used approaches are master equations (e.g.~\cite{Wilson-Rae2008Sep}) or Langevin equations (e.g.~\cite{Genes2008Mar, Bowen2015Nov}). 
Modeling of the Brownian motion of a mechanical resonator is a complex problem, as highlighted in~\cite{Giovannetti2001Jan, Barchielli2015Aug} for instance, and
these two approaches usually do it differently which leads to different values for $\Neff$. Nevertheless, these differences vanish in the limit of  high temperature $T_\m$ and high quality factor $Q_\m$, which is relevant for the setups studied in this article (see Tab.~\ref{tab:Params}). A detailed derivation of both the master equation and the Langevin equations valid beyond the high temperature regime for the mechanical bath is provided  in Ref.~\cite{Barchielli2015Aug}. Typically, in the master equation approach, the Brownian motion is modeled by the dissipative term $\gamma (1+\Nm )  D[\hat{b}]\hat{\rho}+ \gamma \Nm  D[\hat{b}^\dagger]\hat{\rho}$ in the evolution of the density operator $\hat{\rho}$ , where $D[\hat{O}]\hat{\rho} = \hat{O}\hat{\rho} \hat{O}^\dagger - \frac{1}{2} \{\hat{O}^\dagger \hat{O}, \hat{\rho}\}$. This term corresponds to symmetrical noise affecting identically the $\hat{q}$ and $\hat{p}$ quadratures. Conversely, in the Langevin equations, the thermal noise typically only acts on $\hat{p}$ (like in Eqs.~\eqref{standard:Langevin}). However, symmetrical noise can be obtained by making a rotating wave approximation (RWA) during the derivation of the Langevin equations (see Chap.~1 in Ref.~\cite{Bowen2015Nov}).
 
\begin{figure}[htb]
    \includegraphics[width=\linewidth]{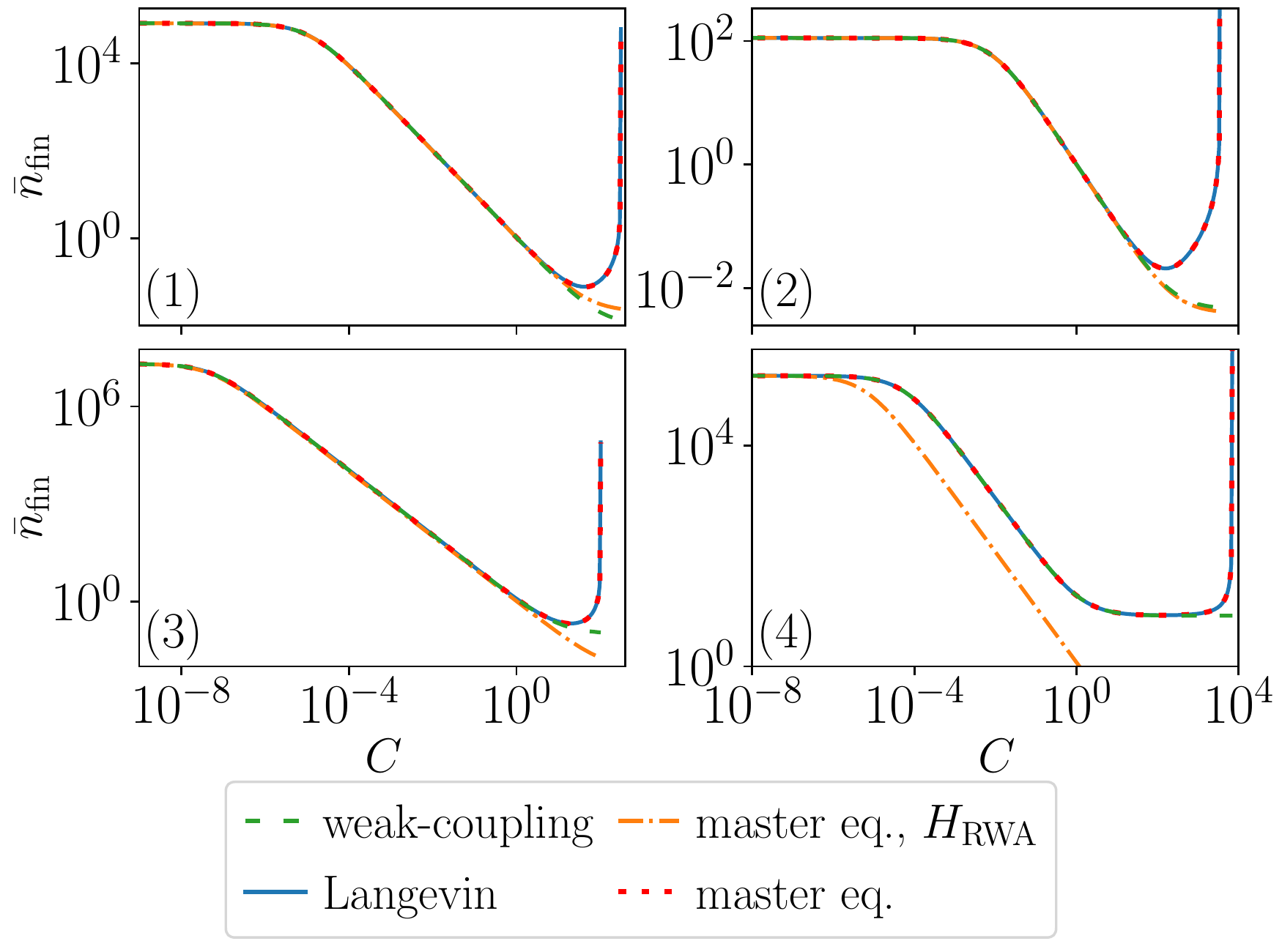}
    \caption{\label{fig:appendix approx comparison}
        Comparison of the resonator's steady-state phonon number $\Neff$ obtain with different methods for each of the systems from Tab.~\ref{tab:Params}.
    }
\end{figure}

Figure~\ref{fig:appendix approx comparison} shows the resonator's steady-state phonon number $\Neff$ computed with different methods: the weak-coupling limit using Eq.~\eqref{weak-coupling:neff} (dashed green line), the master equation approach from Ref.~\cite{Wilson-Rae2008Sep} (dotted red line), the same master equation approach but making an extra RWA (yellow dash-dotted line) and a Langevin approach, going beyond the white-noise approximation for the mechanical noise like in Appendix~\ref{appendix:standard:beyond white noise} and Ref.~\cite{Genes2008Mar} (solid blue line). 
We can see that the master equation and Langevin approaches give almost identical results for the experimental systems considered in this article. Note that the approach followed in the main text to obtain, among others, the plots of $\Neff$ in Fig.~\ref{fig:neff} uses an intermediate level of approximation: Langevin equations with noise on the resonator's momentum only (more accurate than the symmetrical noise in the master equation) but using the white-noise approximation for the correlations of the mechanical noise. 

The green and yellow curves show us at which cooperativity the weak-coupling approximation and RWA respectively stop working. The RWA in question here consists in neglecting the non-resonant terms $\da\db$ and $\da^\dagger\db^\dagger$, and therefore leads to the Hamiltonian
\begin{equation}
    \hat{H}_\text{RWA}=\hbar{\Delta} \da^\dagger \da +\hbar \Om \db^\dagger \db\nonumber-\hbar g (\da^\dagger\db+\da\db^\dagger),
\end{equation}
which  corresponds to an intuitive scattering picture since the total number of particles is conserved.

\section{Squeezed-light setup}\label{appendix:squeezing}

\subsection{Squeezed noise}\label{appendix:squeezing:noise}
    
The correlation functions for the squeezed noise generated by the setup described in Sec.~\ref{sec:II:squeezed-light setup} read~\cite{Asjad2016Nov, Gardiner2004}
\begin{subequations}\label{correl_squeezing}
    \begin{align}
        \mean{\hat{a}^{\dagger}_\text{in}(t)\hat{a}_\text{in}(t')} &= \pi_\s  n_\s(t - t'),\\
        \mean{\hat{a}_\text{in}(t)\hat{a}_\text{in}(t')} &= \pi_\s  m_\s(t - t').
    \end{align}
\end{subequations}
We have defined
\begin{subequations}\label{squeeezing:ns and ms}
    \begin{align}
    n_\s(\tau) &= \frac{r_+^2 - r_-^2}{4}\left(\frac{\e^{-r_-\abs{\tau}}}{2r_-} - \frac{\e^{-r_+\abs{\tau}}}{2r_+}\right), \label{squeezing:ms}\\
    m_\s(\tau) &= \frac{r_+^2 - r_-^2}{4}\left(\frac{\e^{-r_-\abs{\tau}}}{2r_-} + \frac{\e^{-r_+\abs{\tau}}}{2r_+}\right)\e^{-i2\theta},\label{squeezing:ns}
     \end{align} 
\end{subequations}
where we have denoted $r_\pm = \varkappa \pm \abs{\chi}$ the decay rates of the fluctuations of the squeezed and anti-squeezed quadratures. We have assumed that we have $\varkappa > \abs{\chi}$. Then, we make the white-noise approximation (\cite{Gardiner2004}, Chap.~10). Namely, we take the limit $r_+, r_- \rightarrow \infty$ in Eqs.~\eqref{squeeezing:ns and ms} to obtain the expressions of $N_\s$ and $M_\s$ (Eqs.~\eqref{squeeezing:Ns and Ms}) used in the noise autocorrelation functions (Eqs.~\eqref{squeezing:correlations_a_in}). Rewriting $r_\pm = \varkappa (1 \pm r_\s)$, where $r_\s$ is the squeezing ratio defined in Sec.~\ref{sec:II:squeezed-light setup}, we see that the white-noise approximation can be made for any value of $r_\s$ in $[0, 1[$ provided that $\varkappa$ is sufficiently large.\\

\subsection{Solving the dynamics}\label{appendix:squeezing:Langevin}
The only differences between the squeezed-light and the standard setup are the correlation functions of the optical noise, given by Eqs.~\eqref{squeezing:correlations_a_in}. As a consequence, the results from Appendix~\ref{appendix:standard:solution Langevin} still hold while there are some small modifications to the evolution of the second-order moments. In particular, the effective mechanical susceptibility is unchanged (Eq.~\eqref{standard:susceptibility}). This shows us that in the weak-coupling limit, the resonator's coupling strength to the effective cold bath, $\Gopt$, is the same while $\Nopt$ can be made smaller than in the standard setup.

\subsubsection{Lyapunov equation}\label{appendix:squeezing:Lyapunov}
Like in Section~\ref{appendix:standard:Lyapunov}, we derive the Lyapunov equation $\dv{V}{t} = A V+V A^T+ B^\text{squ}$ for the covariance matrix $V$ of the quadratures. $A$ is unchanged (Eq.~\eqref{standard:Lyapunov:A}) while $B^\text{squ} = B + \delta B$, where $B$ is the matrix for the usual vacuum noise, given by Eq.~\eqref{standard:Lyapunov:B} and $\delta B$ the additional term coming from the squeezing:
\begin{equation}
    \!\!\!\frac{\delta B}{2\kappa \pi_\s} = \begin{bmatrix}
        \!N_\s + \abs{M_\s}\cos(2\theta)\!\! & -\abs{M_\s}\sin(2\theta) &\! 0 &\! 0  \\ 
        -\abs{M_\s}\sin(2\theta)& \!\!N_\s - \abs{M_\s}\cos(2\theta)  &\! 0 &\! 0  \\ 
        0& 0 &\! 0 &\! 0 \\
        0 & 0 &\! 0& \!0
    \end{bmatrix}\!\!.\!\!\! \vspace{0.01cm}
\end{equation}
\\

As previously, we determine the steady state from the Lyapunov equation and compute $\Neff$ from Eq.~\eqref{neff_Lyapunov} and the efficiencies $\etaC$ and $\etaL$ from Eqs.~\eqref{etaC_Lyapunov} and Eq.~\eqref{etaL_Lyapunov}. We obtain the following analytical expressions
\begin{align}
    \!\!\Neff &= \frac{g^2 (c_1 + \delta c_1) + \gamma\Nm c_2}{c_3  S_\text{RH}},\\
    \!\!\etaL &= \frac{(4\Om g_{0}^{2} {\gamma} {\kappa}^{2})(c_4 + \delta c_4 + \Nm c_5)}{{\left({\Delta}^{2} + {\kappa}^{2}\right)} \omL c_3},\\
    \!\! \etaC &= \frac{\gamma  S_\text{RH} (c_4 + \delta c_4 + \Nm c_5)}{\Om(c_6 + \delta c_6 + \gamma\Nm c_7)},\!\!
\end{align}
where $c_1,\,c_2, \,c_3, \,c_4, \,c_5, \,c_6$ and $c_7$ are given by Eqs.~\eqref{standard:full analytical expr} and  $S_\text{RH}$ is defined by Eq.~\eqref{standard:stability criterion}. We have defined the corrections coming from the squeezing
\begin{widetext}
    \begin{subequations}
    \begin{align}
       \delta c_1 =\,
        &  2\kappa \pi_\s  N_\s \Big[ \left( \Om ({\kappa}^{2} + \Delta^2)^2 + (\Delta^2 + 2\Om^2)S_\text{RH} \right)\kappa \nonumber \\
       &\hspace{1.2cm} +\gamma \left( \Om\kappa^2(2\kappa^2 + 2\Delta^2 + \Om^2)  + (\Delta^2 + \kappa^2)S_\text{RH} + \Delta\Om^2(\Delta\Om - 2g^2)\right)+\gamma^2 {\left({\Delta}^{2} + {\kappa}^{2}\right)} \Om {\kappa}\Big]\nonumber\\
        &
        -2\kappa \pi_\s \abs{M_\s} \left(c^*\cos(2\theta) + s^*\sin(2\theta)\right),\\
        \delta c_4 =\,
        & 2 \pi_\s \abs{M_\s} \Big[{\left({\Delta}^{2} - \Om^{2} - {\kappa}^{2}- {\gamma} {\kappa} \right)} {\cos(2\theta)} + \Delta{\left( {\gamma} + 2  {\kappa}\right)} {\sin(2\theta)}\Big]\kappa   - 2 \pi_\s N_\s \Big[ ({\Delta}^{2} +  \Om^{2}  +{\kappa}^{2})\kappa +  \gamma{\left({\Delta}^{2} + {\kappa}^{2}\right)} \Big] \nonumber\\
        =\,&\delta c_6,
    \end{align}
    \end{subequations}
    with 
    \begin{subequations}\label{squeezing:optimal angle}
        \begin{align}
            s^* = 
            & \Big( 2  {\left[\Om({\Delta}^{2} +  {\kappa}^{2})+ S_\text{RH}\right]} {\kappa} + \gamma \left[ 5\Om(4  {\kappa}^{2} + \Om^{2}) + S_\text{RH}\right]  +2 \gamma^2 \Om {\kappa} \Big){\Delta} {\kappa},\\
            c^* =\,
            &  \Big[ \Om({\Delta}^{4} - {\kappa}^{4}) + {\left({\Delta}^{2}  - 2 \, {\Om}^{2}  - {\kappa}^{2}\right)} S_\text{RH} \Big]\kappa  + \gamma\Big[ - 3 \Om {\kappa}^{4} + {\left({\Delta}^{2} \Om - \Om^{3} + 4  {\Delta} g^{2}\right)} {\kappa}^{2} + 2  {\Delta} \Om^{2} g^{2}\Big] \nonumber\\
            &+ \gamma^2{\left({\Delta} + {\kappa}\right)} {\left({\Delta} - {\kappa}\right)} \Om {\kappa}.
        \end{align}
    \end{subequations}
\end{widetext}
 We can see that the stability of the optomechanical system is still given by the condition $S_\text{RH} > 0$.

\subsubsection{Parameter optimization}\label{appendix:squeezing:optimization}
The optimal squeezing angle $\theta^*$ is obtained by finding the minimum of $\Neff(\theta)$. We find 
$\cos(2\theta^*) = c^*/\sqrt{c^{*2} + s^{*2}}$ and $\sin(2\theta^*) = s^*/\sqrt{c^{*2} + s^{*2}}$, where $s^*$ and $c^*$ are defined in Eqs.~\eqref{squeezing:optimal angle}. Therefore the optimal squeezing angle depends only on the optomechanical parameters and not on the other squeezing parameters.

We also minimize $\Neff(r_\s)$ to find the optimal squeezing ratio $r_\s^*$. We find that $r_\s^*$ is a root of the fourth degree equation
\begin{equation}
    a^* r_\s^4 - 4b^* r_\s^3+ 6a^*r_\s^2 -4 b^* r_\s +  a^* = 0,
\end{equation} 
with
\begin{subequations}
    \begin{align}
       a^* =&\, c^*\cos(2\theta) + s^*\sin(2\theta)\\
       b^*=
       &\Big[\Om ({\kappa}^{2} + \Delta^2)^2+ S_\text{RH}(\Delta^2 + \kappa^2 + 2\Om^2) \Big]\kappa \nonumber\\
       &+\gamma \Big[ \Om(\kappa^2 + \Delta^2) (2\kappa^2 + \Delta^2) +(\Om^2 + \kappa^2)S_\text{RH}\nonumber\\
       &\hspace{0.75cm}   - 2\Delta^3g^2\Big]+\gamma^2{\left({\Delta}^{2} + {\kappa}^{2}\right)} \Om {\kappa}
    \end{align}
\end{subequations}
The only root that fulfills $0 \le r_\s < 1$ is 
\begin{align}
    r_\s^* =\,&  \frac{b^*}{a^*} + \sqrt{\left(\frac{b^*}{a^*}\right)^{2} - 1}\nonumber\\
    &-\sqrt{2} \sqrt{\left(\frac{b^*}{a^*}\right)^{2} +\frac{b^*}{a^*} \sqrt{\left(\frac{b^*}{a^*}\right)^{2} - 1}  - 1},
\end{align}
which depends on both the optomechanical parameters and the squeezing angle.\\

\subsubsection{Beyond the white-noise approximation}\label{appendix:squeezing:beyond}
Like for the standard setup in Appendix~\ref{appendix:standard:beyond white noise}, we can go beyond the white-noise approximation, but this time for both noises. The correlation functions for the optical noise, in the frequency domain, read
\begin{subequations}
    \begin{align}
        \mean{(\hat{a}^{\dagger}_\text{in})[\omega]\hat{a}_\text{in}[\omega']} &= \pi_\s n_\s[\omega] 2\pi\delta(\omega + \omega'),\\
        \mean{\hat{a}_\text{in}[\omega]\hat{a}_\text{in}[\omega']} &= \pi_\s m_\s[\omega] 2\pi\delta(\omega + \omega'),
    \end{align}
\end{subequations}
with
\begin{subequations}
    \begin{align}
        n_\s[\omega] &= \frac{r_+^2 - r_-^2}{4} \left(\frac{1}{r_-^2 + \omega^2} - \frac{1}{r_+^2 + \omega^2} \right),\\
        m_\s[\omega] &= \frac{r_+^2 - r_-^2}{4} \left(\frac{1}{r_-^2 + \omega^2} + \frac{1}{r_+^2 + \omega^2} \right)\e^{-i2\theta}.
    \end{align}
\end{subequations}

As previously, the steady-state phonon number $\Neff$ can be computed using the position fluctuation spectrum
\begin{equation}
    S_q[\omega] = \abs{\chi_\m^\eff[\omega] }^2(S_\text{th}[\omega] + S^\text{squ}_\text{rp}[\omega]).\label{squeezing:Sq}
\end{equation}
$S_\text{th}[\omega]$ is given by Eq.~\eqref{Sth} and the radiation pressure noise spectrum reads
\begin{widetext}
    \begin{align}
        \hspace*{-0.25cm}S^\text{squ}_\text{rp}[\omega] =\,& S_\text{rp}[\omega]\left(1 +2\pi_\s\frac{ n_\s[\omega] (\kappa^2 + \Delta^2 + \omega^2) - \abs{m_\s[\omega]}\left[( \Delta^2 - \kappa^2 - \omega^2)\cos(2\theta) + 2\Delta\kappa\sin(2\theta)\right]}{\kappa^2 + (\omega + \Delta)^2}\right)\!\!\!\!
    \end{align}
\end{widetext}
where $S_\text{rp}[\omega]$ is the radiation pressure noise spectrum for unsqueezed vacuum noise (Eq.~\eqref{Srp}).

\section{Fano-mirror setup}\label{appendix:Fano}

\subsection{Steady state and linearization}\label{appendix:Fano:lin}

We determine the semi-classical steady state of Eqs.~\eqref{Fano:Langevin} and derive the Langevin equations~\eqref{Fano:Langevin lin} like for the standard setup in Appendix~\ref{appendix:standard:lin}. The difference is the presence of the Fano-mirror mode, whose annihilation operator is split into $\hat{d} =\delta + \deltd$. The steady state $(\alpha,  \bar{q}, \delta)$ is given by the system of non-linear equations~\eqref{Fano:average values}.\\

\begin{figure}[htb]
    \includegraphics[width=\linewidth]{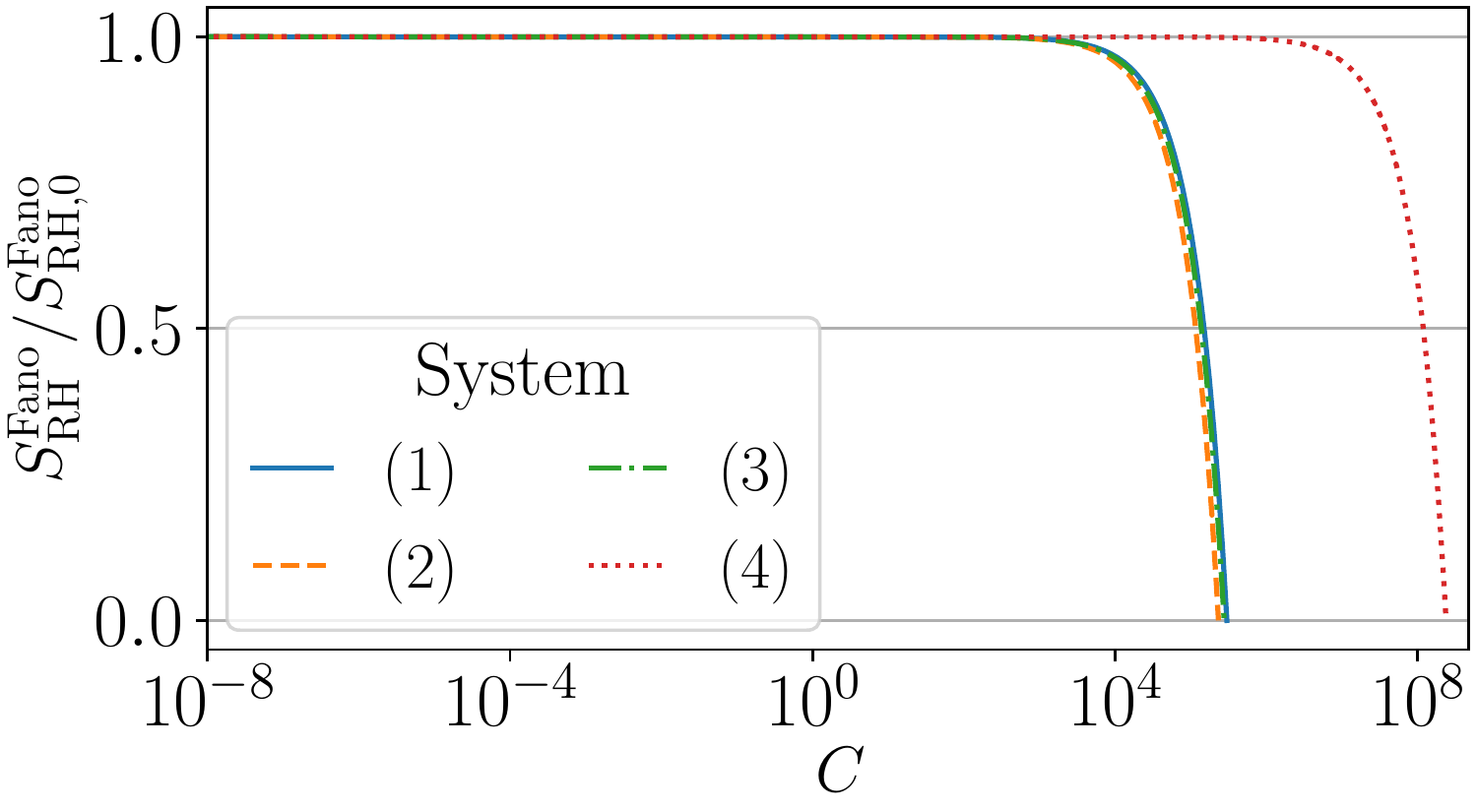}
    \caption{\label{fig:appendix stability Fano}
         Stability condition $S_\text{RH}^\text{Fano}$ of the Fano-mirror setup as a function of the cooperativities considered in this article for each of the systems from Tab.~\ref{tab:Params}. $S_\text{RH}^\text{Fano}$ has been normalized by its value at zero coupling $S_{\text{RH}, 0}^\text{Fano}$.
    }
\end{figure}

In this case, the stability condition is $S_\text{RH}^\text{Fano} > 0$, with 
\begin{align}
    S_\text{RH}^\text{Fano} =\,& {\Delta}^{2} \Om {\gamma_d}^{2} + 4  \Om {\gamma_d}^{2} ({\kappa_0}^{2} -  {\Delta}   \sqrt{{\kappa_0 \kappa_\L}} + {\kappa_0} {\kappa_\L}) \nonumber\\
    &  + 2   \Om {\gamma_d}(\kappa_\L - {\kappa_0}) {\Delta_d}({\Delta} + 2  \sqrt{\kappa_0{\kappa_\L}} ) \nonumber\\
    & +  \Om {\Delta_d}^{2}\left({\Delta}^{2} +  ( {\kappa_0} + {\kappa_\L})^2\right) 
     \nonumber\\
    & - 4  \big( {\gamma_d}^{2}(\Delta - 2 \sqrt{{\kappa_0}{\kappa_\L}})+ {\gamma_d} ({\kappa_\L} - {\kappa_0}) {\Delta_d}  \nonumber\\
        & \hspace{0.75cm}+ {\Delta} {\Delta_d}^{2}\big) g^{2}.
\end{align}
$S_\text{RH}^\text{Fano} > 0$ corresponds, up to a factor $\Om$, to the determinant of the $A$ matrix (Eq.~\eqref{A_Fano}) of the Lyapunov equation (see Section \ref{appendix:Fano:Lyapunov}). 
As can be seen in Fig.~\ref{fig:appendix stability Fano}, the Fano-mirror setup for the four systems studied in this article fulfill this stability condition for all the considered values of the coupling $g$.

\subsection{Solving the dynamics}\label{appendix:Fano:Langevin}

\subsubsection{Solution of the Langevin equations}
Like in Appendix~\ref{appendix:standard:solution Langevin}, the Langevin equations~\eqref{Fano:Langevin lin} can be solved in the frequency domain and the expression of the mechanical position is given by
\begin{align} \label{Fano:solution Langevin}
    \chi_\m^\eff[\omega]^{-1}\dq[\omega] = \;
    & g \sum_{\mu = \L,\R} \Big\{(t_\mu[\omega] + t^*_\mu[-\omega])\hat{X}_{\text{in},\mu}[\omega] \nonumber \\
    & \hspace{1.25cm}+i(t_\mu[\omega] - t^*_\mu[-\omega])\hat{P}_{\text{in},\mu}[\omega] \Big\}\nonumber \\
    &+ \sqrt{\gamma}\hat{\xi}[\omega], 
\end{align}
with the input noise quadratures
$\hat{X}_{\text{in},\mu} = (\hat{a}_{\text{in},\mu} + \hat{a}^\dagger_{\text{in},\mu})/\sqrt{2}$ and $\hat{P}_{\text{in},\mu} = i(\hat{a}^\dagger_{\text{in},\mu} - \hat{a}_{\text{in},\mu}) /\sqrt{2}$.
We have defined $t_\mu[\omega]$ as
\begin{align}
    t_\L[\omega] &= \frac{\sqrt{2\kappa_\L} - \sqrt{2\gamma_d}\mathcal{G}\varepsilon_d[\omega]^{-1}}{\varepsilon_a[\omega] - \mathcal{G}^2\varepsilon_d[\omega]^{-1}},\\
    t_\R[\omega] &= \frac{\sqrt{2\kappa_0} }{\varepsilon_a[\omega] - \mathcal{G}^2\varepsilon_d[\omega]^{-1}},
\end{align}
with $\varepsilon_a[\omega] = \kappa + i(\Delta - \omega)$ and $\varepsilon_d[\omega] = \gamma_d + i(\Delta_d - \omega)$. Note that in this section the $*$ exponent denotes the complex conjugate and not an optimal value like in the part on the squeezing.
The mechanical susceptibility is given by 
\begin{equation}
    \chi_\m^\eff[\omega]^{-1} = \chi^0_\m[\omega]^{-1} + \chi^\eff_\text{opt}[\omega]^{-1},
\end{equation}
where $\chi^0_\m[\omega] = \Om\left(\Om^2 - \omega^2 - i\omega\gamma\right)^{-1}$ is the mechanical susceptibility of the bare resonator and $\chi^\eff_\text{opt}[\omega]$ the optical contribution to the susceptibility, which reads
\begin{equation}
   \chi^\eff_\text{opt}[\omega]^{-1} =  \frac{2ig^2 }{\varepsilon_a^*[-\omega] - \mathcal{G}^{*2}\varepsilon_d^*[-\omega]^{-1}} -\frac{ 2ig^2 }{\varepsilon_a[\omega] - \mathcal{G}^2\varepsilon_d[\omega]^{-1}}.
\end{equation}
We then identify the optical contribution to the mechanical damping rate,
\begin{equation}
    \Gopt[\omega] = -\frac{\Om}{\omega} \Im{\chi^\eff_\text{opt}[\omega]^{-1}}.
\end{equation}
Therefore, in the weak-coupling regime, the resonator's coupling to the effective cold bath, given by $\Gopt[\Om]$, is different from the one of the standard setup.

\subsubsection{Lyapunov equation}\label{appendix:Fano:Lyapunov}
Like in Section~\ref{appendix:standard:Lyapunov}, we write the Langevin equations for the quadratures and obtain the Lyapunov equation $\dv{V}{t} = A V+V A^T+ B$ for $\vec{Y} = \left( \dXa, \dPa, \dq, \deltp, \dXd, \dPd  \right)$, with
{ \footnotesize   \noindent
   \begin{subequations}
        \begin{align}
       \!\! \!\!\!  A &=\! \begin{bmatrix}
            -{\kappa} &\! {\Delta} &\! 0 &\! 0 &\!\!-\sqrt{{\gamma_d}{\kappa_\L}} &\! \sqrt{{\gamma_d}{\kappa_0}} \\
            -{\Delta} &\! -{\kappa} &\! 2  g &\! 0 &\! \!-\sqrt{{\gamma_d}{\kappa_0}} &\! -\sqrt{{\gamma_d}{\kappa_\L}} \\
            0 &\! 0 &\! 0 &\! \Om &\! 0 &\! 0 \\
            2  g &\! 0 &\!\! -\Om &\! -{\gamma} &\! 0 &\! 0 \\
            -\sqrt{{\gamma_d}{\kappa_\L}} &\! \sqrt{{\gamma_d}{\kappa_0}} &\! 0 &\! 0 &\!-{\gamma_d} &\! {\Delta_d} \\
            -\sqrt{{\gamma_d}{\kappa_0}} &\! -\sqrt{{\gamma_d}{\kappa_\L}} &\! 0 &\! 0 &\! -{\Delta_d} &\! -{\gamma_d}
        \end{bmatrix}\label{A_Fano}\!,\!\!\\
        \!\!\! \! \! B &=\!\begin{bmatrix}
            {\kappa} & 0 & 0 & 0 &\!\sqrt{{\gamma_d}{\kappa_\L}} \!& 0 \\
            0 & {\kappa} & 0 & 0 & 0 & \!\sqrt{{\gamma_d}{\kappa_\L}}\! \\
            0 & 0 & 0 & 0 & 0 & 0 \\
            0 & 0 & 0 & {\left(2 {\Nm} + 1\right)} {\gamma} \!& 0 & 0 \\
            \sqrt{{\gamma_d}{\kappa_\L}} & 0 & 0 & 0 & {\gamma_d} & 0 \\
            0 & \sqrt{{\gamma_d}{\kappa_\L}} & 0 & 0 & 0 & {\gamma_d}
        \end{bmatrix}\!.\!\!
    \end{align}
   \end{subequations}
}
These results are in agreement with the ones from the supplementary material of~\cite{Cernotik2019Jun}.
The matrices are now too big to get a tractable analytical solution so we solve the Lyapunov equation numerically instead
(using python and in particular \verb|scipy.linalg.solve_continuous_lyapunov(A, -B)|). As previously, $\Neff$ is given by Eq.~\eqref{neff_Lyapunov} and the efficiencies $\etaC$ and $\etaL$ by Eqs.~\eqref{etaC_Lyapunov} and~\eqref{etaL_Lyapunov}. $\etaC'$ (Eq.~\eqref{etaC_Fano}) is obtained from
\begin{equation}
    \etaC' = \frac{-2 g\bar{V}_{14}}{2g\bar{V}_{23}- \sqrt{\gamma_d\kappa_\L}(\bar{V}_{15} + \bar{V}_{26})+\sqrt{\gamma_d\kappa_0}(\bar{V}_{16} - \bar{V}_{25})}.
\end{equation}

\subsubsection{Beyond the white-noise approximation}\label{appendix:Fano:beyond}
Like for the standard setup in Appendix~\ref{appendix:standard:beyond white noise}, we can go beyond the white-noise approximation for the mechanical noise and compute the effective phonon number from position fluctuation spectrum. Using Eq.~\eqref{Fano:solution Langevin}, we obtain 
\begin{equation}
    S_q[\omega] = \abs{\chi_\m^\eff[\omega] }^2(S_\text{th}[\omega] + S^\text{Fano}_\text{rp}[\omega])\label{Fano:Sq}
\end{equation}
where $S_\text{th}[\omega]$ is given by Eq.~\eqref{Sth} and the radiation pressure noise spectrum reads
\begin{equation}
    S^\text{Fano}_\text{rp}[\omega] = 2g^2(\abs{t_\L[\omega]}^2+\abs{t_\R[\omega]}^2).\label{Fano:Srp}
\end{equation}

\subsubsection{Weak-coupling limit}\label{appendix:Fano:weak-coupling}

We study here the weak-coupling limit $g \ll \kappa, \Om$. In this limit, illustrated in Fig.~\ref{fig:Fano_and_squeezing}(c), the Stokes and anti-Stokes rates are given by 
\begin{equation}
    A^\text{Fano}_\pm = g^2(\abs{t_\L[\mp\Om]}^2+\abs{t_\R[\mp\Om]}^2). \label{Fano:Stokes rates}
\end{equation}
Therefore, at $\Delta_d = \Om$ and the optimal $\gamma_d$, which is relevant for the sideband-resolved systems (1)-(3) considered in this article (see \Tref{tab:Params}), we have
\begin{align}
   \!\! \frac{A^\text{Fano}_-}{A_-} &= \frac{(4\Om)^2}{\Gamma\kappa_\eff},\\
    \!\!\frac{A^\text{Fano}_+}{A_+} &= \frac{2\Gamma\kappa_\eff(4\Om)^2(4\Om^2 + \kappa_\eff^2)}{(4\Om)^2\left((4\Om)^2 + 2\Gamma\kappa_\eff\right)^2 + \Gamma^2\kappa_\eff^4},
\end{align}
where $A_\pm$ are the rates (Eqs.~\eqref{Stokes_rates_perturbative}) for the cavity in the standard setup of equivalent linewidth and associated loss rate $\kappa_\eff$ (see Sec.~\ref{sec:II:Fano-mirror setup}). Namely, in systems (1) to (3), the use of the Fano-mirror setup reduces the rate of the detrimental Stokes process by 10 to 12 orders of magnitude while the rate of the anti-Stokes process that cools down the resonator is only decreased by 2 to 4 orders of magnitude. As a consequence, the temperature of the effective cold phonon bath is lower ($\Nopt$ is 8 to 10 orders of magnitude smaller) but $\Gopt$ is also a couple of orders of magnitude smaller. This explains the shift of curves for the Fano-mirror setup in terms of cooperativity in Fig.~\ref{fig:neff}. Indeed, expressing $\Gopt$ as a function of the cooperativity, $\Gopt = \alpha_\opt C$, and neglecting $\Gopt\Nopt$ in Eq.~\eqref{weak-coupling:neff}, we obtain $\Neff \simeq \gamma \Nm /(\gamma + \alpha_\opt C)$. The reduction of the phonon number starts to be non-negligible when $C$ becomes larger than $\gamma / \alpha_\opt$, but $\alpha_\opt$ is a few orders of magnitude smaller for the Fano-mirror setup, hence the shift in cooperativity. For a system in the resolved-sideband regime, with $\Delta = \Om$, the cooperativity threshold to start seeing a decrease in $\Neff$ is $C \sim 1/\Nm$ for the standard setup and $C \sim \Gamma\kappa_\eff /((4\Om)^2\Nm)$ for the Fano-mirror setup.

\subsubsection{Optimal detuning in the non-resolved-sideband regime}\label{appendix:Fano:non-sideband-resolved}

In the non-sideband-resolved regime, for the standard and squeezed-light setups, the best cooling is not obtained for a detuning $\Delta$ close to $\kappa$. However, for the Fano-mirror setup, the optimal choice is $\Delta_d \simeq 2.5\Om$, not $\Delta_d = \kappa_\eff$. We can understand it by looking at the phonon number in the cold bath in the weak-coupling regime: 
$\Nopt^\text{Fano} = A^\text{Fano}_+/(A^\text{Fano}_- - A^\text{Fano}_+)$ (see Eq.~\eqref{Fano:Stokes rates}). Indeed, the parameters of the Fano-mirror setup are such that $\gamma_d = 4\Om\zeta_0$ and $\zeta_0 = 4\Om / \kappa_\eff$ (see Sec.~\ref{sec:II:experimental setups:Fano}), so in the limit $\Om \ll \kappa_\eff$ and $\gamma_d \ll \Gamma$, we find that the minimum of $\Nopt^\text{Fano}$ is reached for 
\begin{align}
    \frac{\Delta_d}{\Om} &= \frac{{\left(3 \, \sqrt{37} + 26\right)}^{\frac{2}{3}} + 2 \, {\left(3 \, \sqrt{37} + 26\right)}^{\frac{1}{3}} + 7}{3 \, {\left(3 \, \sqrt{37} + 26\right)}^{\frac{1}{3}}} \nonumber\\
    &\simeq 2.5.\\\nonumber
\end{align}

\section{Comparison with absorption refrigerators} \label{appendix:comp absorption refrigerator}
Optomechanical cooling has also been studied in the context of absorption refrigerators \cite{mitchison_realising_2016, Naseem2020Jun}, which are, like the setups we study, autonomous thermal machines.
However, this framework is different from the situation we consider in this article as it typically assumes a three-body interaction, where three system parts interact with three heat baths \cite{Mitchison2018}. One of the baths has to be hotter than the other and can, for example, be realized by thermal light. 

Conversely, in our case, we study a cavity optomechanical system, which is a two-body system (cavity and mechanical resonator), where each subsystem is coupled to a bath. Those baths are typically at the same temperature. Furthermore, we analyze coherent or squeezed light incident to the cavity - which are both minimum uncertainty states - and therefore interpreted as work. The valve picture depicted in Fig.~\ref{fig:principle}(a) is quite intuitive to describe the standard optomechanical setup and can be straightforwardly adapted to the squeezed-light and Fano-mirror setups. In contrast, the result of a quantum absorption refrigerator is that one can cool with thermal light, which we do not consider in our work.

The first setup studied in Ref.~\cite{mitchison_realising_2016} couples the cavity mode to the motion of a trapped atom (mechanical degree of freedom) and to its internal state (two-level system) while the second couples the modes of two cavities to the motion of a trapped atom. Both cases are quite different from the standard optomechanical sideband cooling setup we study and therefore hard to compare. We can nevertheless note that their predictions using sunlight (Fig.~7 in Ref.~\cite{mitchison_realising_2016}) brings the mechanics close to the ground state but $\Neff$ does not go well below 1 as in the levitated optomechanical system which is the most similar system we study (see system (3) in Fig.~\ref{fig:neff} and Ref.~\cite{Delic2020} for the experimental results).

The setup studied in Ref.~\cite{Naseem2020Jun} would correspond to our standard setup where the coherent laser drive has been replaced by thermal light at temperature $T_w$. But their results are hard to compare to ours as they do not study the final phonon number and renormalize their efficiency by the Carnot absorption refrigerator efficiency, which is ill-defined in our two-terminal system with possibly equal temperatures.

\bibliography{main.bib}

\end{document}